\documentclass[12pt]{article}
\usepackage[margin=1in]{geometry}
\usepackage[utf8]{inputenc}
\usepackage{amsmath}
\usepackage{amsfonts}
\usepackage{bm}
\usepackage{graphicx, epsfig}
\usepackage{xcolor}
\usepackage{setspace}
\usepackage{enumerate}
\usepackage{graphicx}
\usepackage{booktabs}
\usepackage{multirow}
\usepackage{booktabs}
\usepackage{comment}
\usepackage{natbib}
\usepackage{tikz}
\newcommand{\pperp}{\perp\!\!\!\perp }
\newcommand{\iid}{\stackrel{\text{iid}}{\sim}}
\newcommand{\Bernoulli}{\operatorname{Bernoulli}}

\definecolor{tred}{HTML}{C136CD}

\newcommand{\lsem}{\textnormal{lsem}}
\newcommand{\bcmf}{\textnormal{bcmf}}

\title{Estimating Heterogeneous Causal Mediation Effects with Bayesian Decision Tree Ensembles}
\author{
    \normalsize Angela Ting and Antonio R. Linero \\
    \normalsize \emph{Department of Statistics and Data Sciences} \\
    \normalsize \emph{University of Texas at Austin}
}

\usepackage[colorlinks,citecolor=blue]{hyperref}
\usepackage{cleveref}

\begin{document}

\maketitle

\begin{abstract}
  The causal inference literature has increasingly recognized that explicitly targeting treatment effect heterogeneity can lead to improved scientific understanding and policy recommendations. Towards the same ends, studying the causal pathway connecting the treatment to the outcome can be also useful. This paper addresses these problems in the context of \emph{causal mediation analysis}. We introduce a varying coefficient model based on Bayesian additive regression trees to identify and regularize heterogeneous causal mediation effects; analogously with linear structural equation models, these effects correspond to covariate-dependent products of coefficients. We show that, even on large datasets with few covariates, LSEMs can produce highly unstable estimates of the conditional average direct and indirect effects, while our \emph{Bayesian causal mediation forests} model produces estimates that are stable. We find that our approach is conservative, with effect estimates ``shrunk towards homogeneity.'' We examine the salient properties of our method using both data from the Medical Expenditure Panel Survey and empirically-grounded simulated data. Finally, we show how our model can be combined with posterior summarization strategies to identify interesting subgroups and interpret the model fit.
\end{abstract}

\doublespacing

\section{Introduction}
\label{sec:introduction}

Estimation of heterogeneous causal effects from observational data is a topic of fundamental importance, with applications in personalized medicine \citep{obermeyer2016predicting}, policy recommendation \citep{athey2017beyond}, and social science \citep{yeager2019national}. A question of great recent interest in the causal inference literature is how best to leverage state-of-the-art prediction algorithms developed in the machine learning community to estimate heterogeneous treatment effects \citep{kunzel2019metalearners, nie2021quasi, hahn2020bayesian}. Much of this literature has focused on the question of how best to modify the estimation strategies used in the machine learning literature to be appropriate for inferring heterogeneous causal effects.

A complementary approach to making better policy recommendations is to learn how a treatment of interest influences the outcome via its effects on downstream variables that are themselves causally linked to the outcome; this is referred to as \emph{causal mediation analysis} and the intermediate variables are referred to as \emph{mediators} \citep{robins1992identifiability, pearl2001direct, rubin2004direct}. In addition to providing a sharper understanding of the causal mechanisms at play, we will see that causal mediation analysis can in some cases increase our power to detect causal effects. Similar questions about how to effectively leverage predictive algorithms have emerged in this field, with much of the focus on estimating \emph{average}, rather than heterogeneous, mediation effects \citep{farbmacher2022causal, linero2022mediation, zheng2012targeted, tchetgen2012semiparametric, kim2017framework}.

To the best of our knowledge, there has been limited work at the intersection of these two settings, i.e., where one is interested in estimating treatment effect heterogeneity at the level of direct and indirect causal mediation effects using machine learning. The issue of estimating heterogeneity in treatment effects in the context of mediation analysis is referred to as \emph{moderated mediation} \citep{muller2005moderation}. This topic has garnered significant attention in the social science literature, often utilizing linear structural equation modeling (LSEM). For example, \citep{preacher2007addressing} and  \citep{kershaw2010socioeconomic} applied moderated mediation using LSEMs to problems in education and health psychology, respectively.

Estimating heterogeneous mediation effects in a nonparametric manner is a challenging task that relies on both strong assumptions regarding confounding and requires large amounts of data to reliably estimate the causal effects. There are important challenges in this context that need to be addressed, including: (i) determining how to properly regularize both the nuisance parameters and parameters of interst to ensure sensible results; (ii) developing methods to summarize the results of black-box fitting procedures in a meaningful way; and (iii) establishing reliable techniques to identify subgroups for which there is evidence of moderated mediation and to determine which variables are acting as effect modifiers.

This paper proposes a two-layer extension of the Bayesian causal forests (BCF) algorithm for estimating heterogeneous mediation effects, which combines a standard BCF model for the mediator with a varying coefficient BART model for the outcome \citep{hahn2020bayesian,deshpande2020vcbart}. Our approach is motivated by the strong performance of BCFs in causal inference competitions and in practical applications \citep{dorie2019automated}. Our approach directly parameterizes the models in terms of the direct and indirect effects of the treatment on the outcome. This allows us to ``shrink towards homogeneity,'' stabilizing the estimation of the mediation effects. Our approach performs extremely well in regimes where treatment effects are nearly homogeneous, with small root-mean squared errors for individual-level mediation effects and credible intervals that attain close to the nominal rate of coverage for most individuals. Hence, our proposed approach provides a powerful tool for estimating heterogeneous mediation effects.



\subsection{The Medical Expenditure Panel Survey}

The Medical Expenditure Panel Survey (MEPS) is an ongoing large-scale survey administered by the Agency for Healthcare Research and Quality that aims to measure the healthcare system's use by patients, hospitals, and insurance companies. To demonstrate our proposed methodology, we employ the MEPS to investigate the health consequences of smoking. Specifically, we aim to answer the following questions: (i) does smoking have a causal impact on healthcare expenses overall? (ii) to what extent is this impact mediated (or not) by smoking's effect on overall health? and (iii) are there any moderating variables that affect the association between smoking and medical expenditures?

In Section~\ref{sec:medical}, we present an analysis of this dataset which yields a surprising finding: the total causal effect of smoking on medical expenditures can be masked by instability resulting from the estimation of the direct effect of smoking on healthcare costs. Although one might intuitively assume that the  effect of smoking on medical expenditures is fully mediated by its impact on health, our analysis under sequential ignorability shows that the estimated direct effect of smoking on expenses is negative and largely counteracts the positive indirect effect of smoking on expenses; this direct effect is likely due to additional variables that we have not incorporated in the analysis. Additionally, we identify several variables, with age being the most important, that moderate the indirect effect of smoking on expenditures.

\subsection{Outline}

In Section~\ref{sec:review} we review the potential outcomes framework for mediation, the sequential ignorability assumption, the Bayesian additive regression trees (BART) framework, and Bayesian causal forests (BCFs). In Section~\ref{sec:bart_mediation} we define our Bayesian causal mediation forests model, and show how to use it to stably estimate the direct and indirect effects. In Section~\ref{sec:medical} we use our methodology to analyze data from the MEPS data to study mediation effect heterogeneity in the effect of smoking on health care expenditures as mediated by the effect of smoking on health, and conduct an empirically-designed simulation study to show that our method performs well in terms of coverage and estimation error for estimating both average and conditional average mediation effects. We conclude in Section~\ref{sec:discussion} with a discussion and possible extensions. Computational detials and further simulation results are given in the supplementary material.



\section{Review of Mediation Analysis and BART}
\label{sec:review}

\subsection{Overview of Mediation Analysis}
\label{sec:overview}

\begin{figure}
  \centering
  \includegraphics[width = 0.3\textwidth]{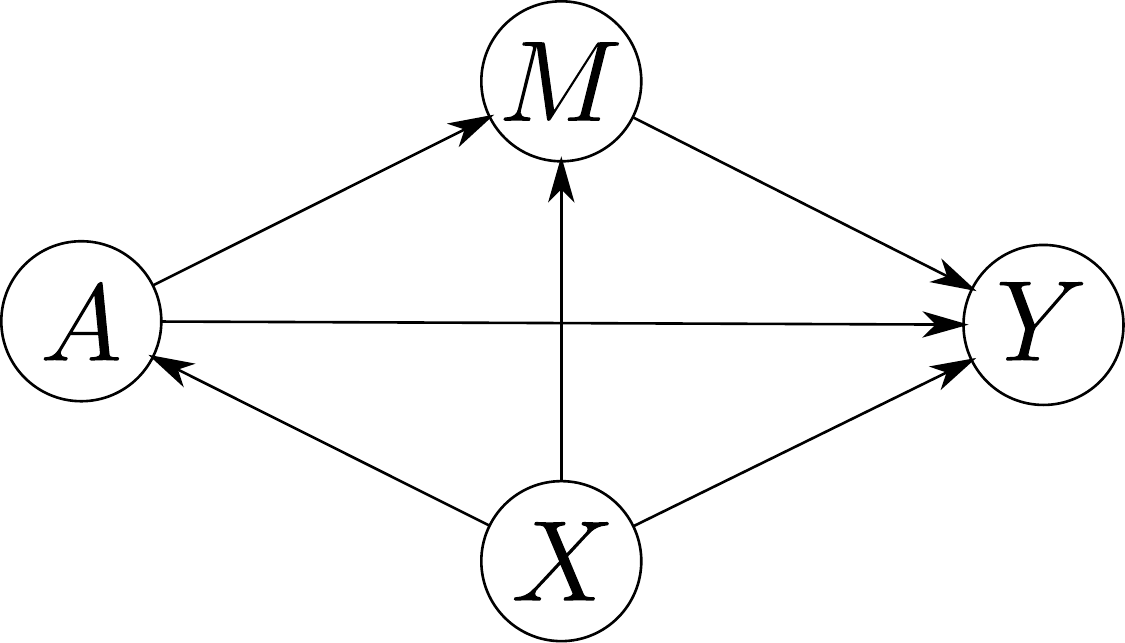}
  \caption{A schematic representation of a treatment $A$, mediator $M$, outcome $Y$, and confounders/effect modifiers $X$. Arrows depict the direction of causality.}
  \label{fig:MedFig}
\end{figure}

Mediation refers to the process through which a treatment ($A$) influences an outcome ($Y$) by acting through an intermediate \emph{mediator} variable ($M$), which occurs between the treatment and the outcome; a graphical representation is given in Figure~\ref{fig:MedFig}. For example, let us consider the question of whether smoking affects medical expenditures directly and indirectly through its effect on health. Here, smoking status is a binary treatment ($A$), and the outcome of interest is the logarithm of medical expenditure ($Y$). Our aim is to break down the effect of smoking on medical expenditures into a \emph{direct effect} of smoking and an \emph{indirect effect} that is mediated by smoking's effect on overall health (measured as an individual's self-perceived quality of health). A natural hypothesis is that smoking does not directly cause higher medical expenditures but rather does so by reducing a person's overall health. Health is on the causal path between the treatment (smoking) and the outcome (medical expenditures) and hence is a mediator.

Mediation analysis has been applied in many scientific fields including epidemiology, medicine, economics, and the social sciences \citep{mackinnon1993estimating, rubin2004direct, mackinnon2008introduction, albert2008mediation, imai2010general, vanderweele2016mediation}. Much of this literature has focused on structural equation models (SEMs) to quantify mediation effects as products of coefficients in parametric models. In particular, linear structural equation models (LSEMs) have been widely used \citep{baron1986moderator, mackinnon1993estimating, mackinnon2008introduction}.

LSEMs have a major limitation in that the identification of the mediation effects is tied to the choice and correct specification of a particular parameteric model, limiting their applicability. To address this limitation, \citet{imai2010general} proposed a nonparametric approach based on \emph{potential outcomes} \citep{rubin2004direct,rubin1974estimating} that allows for the identification of average causal mediation effects under the assumption of \emph{sequential ignorability}. This assumption states that the treatment is independent of all potential values of the outcome and mediator given the covariates, and the observed mediator is independent of all potential outcomes given the observed treatment and covariates. By avoiding parametric assumptions, this framework provides a general estimation procedure that is agnostic to the choice of model for the outcome and mediator, making it applicable in a wide range of settings. 


For individuals $i = 1,...,n$ and treatment $a \in \{0,1\}$, define the potential outcome $M_i(a)$ as the value of the mediator that would have been observed had the individual received treatment $a$. Note that for each individual, only one of $M_i(0)$ or $M_i(1)$ is actually observed. For treated individuals ($A_i = 1$), $M_i(0)$ is a \emph{counterfactual}, i.e., the value of the mediator that would have been observed had the individual been untreated instead. Similarly, the potential outcome $Y_i(a,m)$ is the value of the outcome that would have been observed had the individual received treatment $a$ and had a mediator at level $m$. For example, $Y_i\{0, M_i(1)\}$ is the value of the outcome that would have been observed if the individual was not treated and had a value of the mediator at the same level they would have had if they were treated. We link the potential outcomes to the observed data through the \emph{consistency} assumption, which states that we observe the mediator $M_i = M_i(A_i)$ and the outcome $Y_i = Y_i\{A_i, M_i(A_i)\}$. Because the values of $Y_i$ and $M_i$ are defined only in terms of the treatment $a$ potentially received by individual $i$ (and not on the treatment received by other individuals), this notation also implicitly states that there is no interference between units, which is known as the \emph{Stable Unit Treatment Value} (SUTVA) assumption.

Using these potential outcomes, we can define the causal estimates of interest. In causal mediation analysis, we are particularly interested in estimating the natural direct and natural indirect effects \citep{pearl2001direct, robins1992identifiability}. The \textit{natural direct effect} is defined as 
\begin{align} \label{eq:nde}
  \zeta_a = E[Y_i\{1, M_i(a)\} - Y_i\{0, M_i(a)\}]
\end{align}
and the \textit{natural indirect effect} is defined as
\begin{align} \label{eq:nie}
  \delta_a = E[Y_i\{a, M_i(1)\} - Y_i\{a, M_i(0)\}].
\end{align}
The natural direct effect isolates the effect of the treatment while keeping the potential mediator fixed, and can be interpreted as the effect that the treatment has directly on the outcome $Y_i$. Conversely, the natural indirect effect isolates the effect of the potential mediator in response to different treatment values while keeping the treatment fixed, and can be interpreted as the effect that the treatment has indirectly on the outcome $Y_i$ through the mediator $M_i$. The total effect of the treatment on the outcome is a sum of the direct and indirect effects, and can be defined as
\begin{align} \label{eq:te}
  \tau
  = \zeta_0 + \delta_1
  = \zeta_1 + \delta_0
  = E[Y_i\{1, M_i(1)\} - Y_i\{0, M_i(0)\}].
\end{align}
We can similarly define \emph{conditional average} variants of both the direct and indirect effects as
\begin{align}
  \label{eq:cate}
  \begin{split}
    \zeta_a(x)
    &=
      E[Y_i\{1, M_i(a)\} - Y_i\{0, M_i(a)\} \mid X_i = x]
      \qquad \text{and} \qquad \\
    \delta_a(x)
    &=
      E[Y_i\{a, M_i(1)\} - Y_i\{a, M_i(0)\} \mid X_i = x].
  \end{split}
\end{align}
Most of our attention will be on the conditional average direct and indirect effects, as defined in \eqref{eq:cate}.

\subsection{Assumptions}
\label{sec:assumptions}

Let the statement $[A \pperp B \mid C = c]$ mean that $A$ is conditionally independent of $B$ given that $C = c$, let $\mathcal X$ denote the sample space of $X_i$, and let $\mathcal M$ denote the sample space of $M_i$. Following \citet{imai2010general}, we make the following \emph{sequential ignorability} assumption throughout, allowing for the identification of the direct and indirect effects.
\begin{description} 
    \item[SI1] $\{Y_i(a', m), M_i(a)\} \pperp A_i \mid X_i = x$ for $a,a' = 0,1$ and all $x \in \mathcal{X}$.
    \item[SI2] $Y_i(a',m) \pperp M_i(a) \mid A_i = a, X_i = x$ for $a,a' = 0,1$ and all $x \in \mathcal{X}$.
    \item[SI3] $\text{Pr}(A_i = a \mid X_i = x) > 0$ and $f\{M_i(a) = m \mid A_i = a, X_i = x\} > 0$ for $a = 0,1$ and all $x \in \mathcal{X}$ and $m \in \mathcal{M}$.
\end{description}
The first assumption states that, given the covariates, the treatment assignment is ignorable, i.e., it is independent of potential outcomes and potential mediators. This assumption is automatically satisfied when individuals are randomly assigned to treatment and control groups, but is not guaranteed to hold in observational studies, in which case researchers often collect as many pre-treatment confounders as possible so that treatment assignment ignorability is plausible after the differences in covariates between treatment groups are accounted for. The second assumption states that, given the observed treatment and covariates, the mediator is ignorable, i.e., it is independent of potential outcomes. This assumption, however, is not guaranteed to hold even in randomized experiments. In general, it cannot be directly tested from the data. The third assumption is a positivity assumption for the treatment and mediator, stating that the probability of receiving the treatment and control should be nonzero.

Under SI1--SI3, we can identify the distribution of any counterfactual outcome $Y_i\{a', M_i(a)\}$ nonparametrically as
\begin{align} \label{eq:nonpar_iden_eq}
  \begin{split}
    &f(Y_i\{a, M_i(a')\} = y \mid X_i = x)
    \\&=
    \int_{\mathcal M} f(Y_i = y \mid M_i = m, A_i = a, X_i = x)
    \, f(M_i = m \mid A_i = a, X_i = x)
    \ dm
  \end{split}
\end{align}
for any $x \in \mathcal{X}$ and $a,a'=0,1$ \citep[][Theorem 1]{imai2010general}. This allows us to make inferences about unobserved counterfactuals (left-hand side) using observed outcomes and mediators (right-hand side). Moreover, \eqref{eq:nonpar_iden_eq} is not dependent on a specific parametric model, and so can be applied to flexible (nonparametric) models.

\subsection{A Review of Bayesian Additive Regression Trees}
\label{sec:bart}

We will use the Bayesian Additive Regression Trees (BART) model proposed by \citet{chipman2010bart}. Consider an unknown function $r$ that predicts an output $Y_i$ using a vector of inputs $X_i$
\begin{align}
  \label{eq:reg_func}
  Y_i = r(X_i) + \epsilon_i, \quad \epsilon_i \sim N(0, \sigma^2).
\end{align}
BART models $r(x)$ as a sum of $m$ regression trees
\begin{math}
  r(x) = \sum_{j=1}^m g(x ; T_j, M_j)
\end{math}
where $T_j$ is a binary decision tree consisting of interior node decision rules as well as a set of terminal nodes and $M_j = \{\mu_{j1}, \ldots, \mu_{jb_j}\}$ is a set of parameter values associated with each of the $b_j$ terminal nodes of tree $T_j$. Each $x$ is associated with a single terminal node $k$ of $T_j$ and is then assigned the value $g(x ; T_j, M_j) = \mu_{jk}$. Under \eqref{eq:reg_func}, $E(Y_i \mid X_i = x)$ equals the sum of all the terminal node $\mu_{jk}$'s assigned to $x$ by the $g(x; T_j, M_j)$'s. For a comprehensive review of BART and its applications, see \citet{hill2020bayesian}.

To apply BART it is necessary to specify a prior distribution over all the parameters of the sum-of-trees model, i.e., $(T_j, M_j)$ for $j = 1,\ldots,m$. This prior should regularize the fit by keeping individual tree effects from being disproportionately influential. The prior consists of two components: a prior for each tree $T_j$ and a prior on the terminal nodes $M_j \mid T_j$ where $\pi(T_j, M_j) = \pi_T(T_j)\pi_M(M_j  \mid  T_j)$. The BART model then sets $(T_j, M_j) \iid \pi(T, M)$.

The prior $\pi_T$ is determined by three variables: (i) the probability that a given node is an interior node, (ii) the distribution of the splitting variable assignments at each interior node, and (iii) the distribution of the splitting rule assignment in each interior node conditional on the splitting variable. For (i), the probability that a node at depth $d$ is an interior node is
\begin{align}
  \label{eq:growtree}
  \alpha(1+d)^{-\beta}, \quad \alpha \in (0,1), \beta \in [0, \infty)
\end{align}
with $\alpha = 0.95$ and $\beta = 2$ being a default that favors small trees. For (ii) and (iii), the distribution of the splitting variable assignments at each interior node and the distribution of the splitting rule assignment in each interior node conditional on the splitting variable are both given a uniform prior.

For the prior on the terminal nodes, we assume $\pi(M_j \mid T_j) = \prod_{k = 1}^{b_j} \pi_\mu(\mu_{jk})$. To specify $\sigma_\mu$, in this paper we first shift and rescale the $Y_i$'s so that $Y_i$ has mean $0$ and variance $1$. We then use the prior
\begin{align*}
  \pi_\mu(\mu_{jk}) = N(\mu_{jk} \mid 0, \sigma^2_\mu)
  \qquad \text{where} \qquad
  \sigma_\mu = \frac{3}{k \sqrt m}
\end{align*}
for a suitable value of $k$, with default $k=2$. Note that this prior shrinks the terminal node values $\mu_{jk}$ towards zero and applies greater shrinkage as the number of trees $m$ is increased, ensuring that each tree is a weak learner in the ensemble of trees.

\subsection{Bayesian Decision Tree Ensembles for Causal Inference}

BART has been seen to perform particularly well in causal inference problems for inferring heterogeneous and average treatment effects \citep{hill2011bayesian, wendling2018comparing, dorie2019automated}. For an outcome $Y_i$, binary treatment $A_i$, and confounder/modifier variables $X_i$, \citet{hill2011bayesian} proposes the model
\begin{align}
  \label{eq:hill}
  Y_i(a) = \mu(X_i, a) + \epsilon_i, \quad \epsilon_i \sim N(0, \sigma^2).
\end{align}
The effect of receiving the treatment is therefore given by
\begin{align*}
  E\{Y_i(1) - Y_i(0) \mid X_i = x\} = \tau(x) = \mu(x, 1) - \mu(x, 0).
\end{align*}
Given BART's strong predictive performance, \citet{hill2011bayesian} suggests using a BART prior for $\mu(\cdot, \cdot)$ to flexibly model the outcome and hence obtain flexible treatment effect estimates.

\citet{hahn2020bayesian} note that successful predictive modeling depends largely on careful regularization, and extend the work of \citet{hill2011bayesian} by noting two shortcomings of the model \eqref{eq:hill}: first, the correlation between the propensity score and $\mu(x,a)$ can induce \emph{regularization induced confounding} (RIC), leading to highly biased causal estimates and, second, priors based on the parameterization \eqref{eq:hill} encode prior information that treatment effects are highly non-homogeneous. To mitigate RIC they develop a prior that depends on an estimate of the propensity score $\widehat\pi_i$ as a 1-dimensional summary of the covariates, while to address non-homogeneity they reparameterize the regression as
\begin{align*}
  Y_i(a) = \mu(X_i, \widehat{\pi}_i) + a \, \tau(X_i) + \epsilon_i
\end{align*}
where $\mu(x,\widehat\pi)$ captures the prognostic effect of the control variables $X_i$ and $\tau(x)$ is exactly the treatment effect. Independent BART priors are then placed on $\mu(\cdot)$ and $\tau(\cdot)$, with the prior on $\tau(\cdot)$ encoding our prior beliefs about the degree of treatment effect heterogeneity.

\citet{linero2022mediation} consider estimation of direct and indirect effects in causal mediation using BART models. Linked with the concept of RIC, they also show that naively specified priors can be highly \emph{dogmatic} \citep{linero2021nonparametric} in the sense of encoding a prior belief that, on average, the mediator and outcome potential outcomes are unconfounded (and hence that inference for the average mediation effects can proceed as though there were no confounding present). Prior dogmatism induces regularization-induced confounding by giving a strong prior preference to encourage the model to attribute causal effects on the outcome as being due to the treatment rather than the confounders. To address this issue, they include ``clever covariates'' $\widehat m_{ai} = \widehat E(M_i \mid A_i = a, X_i)$ into the outcome model for $a \in \{0,1\}$. These clever covariates are analagous to the propensity score estimate $\widehat\pi_i$ used to correct for RIC in the BCF. \citet{linero2022mediation} then introduce the \emph{Bayesian causal mediation forests} (BCMF) model
\begin{align}
  \label{eq:ogbcmf}
  \begin{split}
  Y_i(a, m) &= \mu_y(m, a, X_i) + \epsilon_i \\
  M_i(a) &= \mu_m(a, X_i) + \epsilon_i
  \end{split}
\end{align}
where the functions $\mu_y(\cdot,0,\cdot)$, $\mu_y(\cdot,1,\cdot)$,
$\mu_m(0,\cdot)$, and $\mu_m(1,\cdot)$ are given independent BART priors and the
clever covariates $\widehat m_{0i}$ and $\widehat m_{1i}$ are included as
predictors into the BART model for the outcome.

While the model \eqref{eq:ogbcmf} accomplishes the goal of estimating average
mediation effects well (i.e., it solves the problem of RIC), it does not
appropriately control the degree of heterogeneity in the conditional average
mediation effects. A contribution of this work is to use the insights behind
the parameterization of BCFs to develop a model that applies seperate regularization to the direct and indirect effects.

\section{BART for Heterogeneous Mediation Effects}
\label{sec:bart_mediation}

We now introduce our causal mediation analysis model; a ``Bayesian backfitting'' algorithm for fitting this model is given in the supplementary material. Analogous to BCFs, the models presented enable direct regularization of $\delta_a(x)$ and $\zeta_a(x)$. This type of direct regularization has been shown to be crucial in generating dependable estimates of heterogneous causal effects in other contexts \citep{hahn2020bayesian, nie2021quasi}.

For numeric outcomes and mediators, we specify the models
\begin{align}
  \label{eq:outcome-bart}
  Y_i(a,m) & = \mu(X_i) + a \, \zeta(X_i) + m \, d(X_i) + \epsilon_i, \\
  \label{eq:mediator-bart}
  M_i(a)   & = \mu_m(X_i) + a \, \tau_m(X_i) + \nu_i
\end{align}
where independent BART priors are specified for $(\mu, \zeta, d, \mu_m, \tau_m)$. The mediator model \eqref{eq:mediator-bart} simply corresponds to a BCF model as proposed by \citet{hahn2020bayesian}, with $\tau_m(x)$ corresponding to a heterogeneous causal effect of the treatment on the outcome. The outcome model \eqref{eq:outcome-bart}, on the other hand, corresponds to a \emph{varying coefficient BART} (VC-BART) model as proposed by \citet{deshpande2020vcbart}, with the treatment $A_i$ and mediator $M_i$ entering linearly.

The model \eqref{eq:outcome-bart}--\eqref{eq:mediator-bart} is a varying coefficient version of commonly used LSEMs, with the coefficients modeled nonparametrically as a function of $X_i$. Because of this, the conditional average mediation effects are also expressible as products of coefficients as
\begin{align*}
  \zeta_a(x)  & = \zeta(x), \qquad\qquad \text{and} \\
  \delta_a(x) & = \tau_m(x) \, d(x).
\end{align*}
Hence, this parameterization allows us to isolate the components $\zeta(x)$ and $\delta(x)$ and apply differing amounts of regularization to them. Note that \eqref{eq:outcome-bart}--\eqref{eq:mediator-bart} assumes that no interaction  exists between the mediator and treatment in the outcome model, and hence $\delta_a(x)$ and $\zeta_a(x)$ do not depend on the treatment level $a$, i.e., $\delta_0(x) = \delta_1(x)$ and $\zeta_0(x) = \zeta_1(x)$.

For average effects, note that the marginal distribution of $Y_i\{a, M_i(a')\}$ is given by 
\begin{align} \label{eq:counterfactual_means}
  f\big(Y_i\{a, M_i(a')\} = y\big)
  = \int f\big(Y_i\{a, M_i(a')\} = y \mid X_i = x \big) \, f(X_i = x) \ dx.
\end{align}
It is therefore necessary to specify a model for the distribution of the covariates. Often, when this distribution is not modeled explicitly, the empirical distribution is used instead as an estimate, i.e. $F_X(dx) = \sum_i \omega_i \, \delta_{X_i}(dx)$ where $\delta_x(\cdot)$ denotes a point-mass distribution at $x$ and $\omega_i = n^{-1}$. An alternative to the empirical distribution is the \emph{Bayesian Bootstrap} \citep[BB][]{rubin1981bayesian}, which respects our inherent uncertainty in $F_X$ while cleanly avoiding the need to model the distribution of the covariates. The BB is similar to the empirical distribution, but instead of setting $\omega_i = n^{-1}$ we use an improper prior $\pi(\omega) = \prod_i \omega_i^{-1}$; this leads to the posterior distribution $\omega \sim \text{Dirichlet}(1,\ldots,1)$ for the weights. Under the BB the average effects are identified as
\begin{math}
  \bar \delta = \sum_i \omega_i \, \delta(X_i)
  \qquad \text{and} \qquad
  \bar \zeta = \sum_i \omega_i \, \zeta(X_i).
\end{math}

\subsection{Controlling Heterogeneity Through Prior Specification}
\label{sec:controlling}

An important advantage of the models \eqref{eq:outcome-bart}--\eqref{eq:mediator-bart} relative to the model of \citet{linero2022mediation} is that we can shrink the model fits towards \emph{homogeneous} mediator and treatment effects through judicious choice of hyperparameters; after doing this, we can be confident that any heterogeneity we \emph{do} detect is well-supported by the data, rather than being the result of instability due to the use of nonparametric estimators.

The degree of heterogeneity of the direct effect can be controlled via the prior specification for $\zeta(x)$. Specifically, we can shrink $\zeta(x)$ to a constant function, with few effect moderators, by (i) setting the parameter $\alpha$ in \eqref{eq:growtree} to a small value (say, $\alpha = 0.5$) so that most trees do not include covariates and (ii) using a smaller number of trees (say, $m = 20$). Using the same strategies, we can control the degree of heterogeneity in $\tau_m(x)$, which represents the causal effect of the treatment on the mediator.

The considerations for the indirect effects are slightly more complicated, as $\delta(x) = d(x) \, \tau_m(x)$ consists of two components. Note that heterogeneity in $\delta(x)$ is inevitable if $\tau_m(x)$ is non-constant. However, if $\tau_m(x)$ is constant then $\delta(x)$ can be made homogeneous by shrinking $d(x)$ towards a constant function. Accordingly, we adopt the same strategy for $d(x)$ as we adopt for $\zeta(x)$ and $\tau_m(x)$: using a small number of trees and setting $\alpha$ small.

\subsection{Modeling Non-Numeric Data}

Binary mediators can also be easily incorporated by using the nonparametric probit regression model
\begin{align*}
  [M_i(a) \mid X_i = x]
  &\sim \Bernoulli[ \Phi\{\mu_m(x) + a \, \tau_m(x)\}],
\end{align*}
with BART priors again used for $(\mu_m, \tau_m)$. Under sequential ignorability and \eqref{eq:outcome-bart}, we can identify the mediation effects as
\begin{align*}
  \zeta_a(x) = \zeta(x)
  \qquad \text{and} \qquad 
  \delta_a(x) = d(x) \left[ \Phi\{\mu_m(x) + \tau_m(x)\} - \Phi\{\mu_m(x)\} \right]
\end{align*}
where $\Phi(\cdot)$ is the cumulative distribution function of a standard normal random variable.
Similar expressions for the direct and indirect effects can also be computed when $[Y_i(a,m) \mid X_i = x] \sim \Bernoulli[\Phi\{\mu(x) + a \, \zeta(x) + m \, d(x)\}]$ and $[M_i(a) \mid X_i = x] \sim N\{\mu_m(x) + a \, \tau_m(x), \sigma^2_m\}$ by noting that
\begin{align*}
  E[Y_i\{a, M_i(a')\} \mid X_i = x]
  =
  \Phi\left( \frac{\mu(x) + a \, \zeta(x) + \{\mu_m(x) + a' \, \tau_m(x)\} \, d(x)}{\sqrt{1 + d^2(x) \, \sigma_m^2}} \right),
\end{align*}
which can be derived by noting that the probit model implies that $E[Y_i\{a, M_i(a')\} \mid X_i = x] = \Pr(\epsilon_i \le \mu(x) + a \, \zeta(x) + M_i(a') \, d(x) \mid X_i = x)$ where $\epsilon_i \sim N(0,1)$ and $\epsilon_i$ is independent of $M_i(a')$. This implies, for example, that when $Y_i$ is binary and $M_i$ is continuous we have
\begin{align*}
  \delta_a(x)
  =
  \Phi\left(
  \frac{\mu(x) + a \, \zeta(x) + d(x) \, \{\mu_m(x) + \tau_m(x)\}}
       {\sqrt{1 + d^2(x) \, \sigma^2_m}} \right)
  -
  \Phi\left( \frac{\mu(x) + a \, \zeta(x) + d(x) \, \mu_m(x)}
                  {\sqrt{1 + d^2(x) \, \sigma^2_m}} \right).
\end{align*}





\subsection{Corrections for Regularization Induced Confounding}

As shown by \citet{linero2022mediation}, Bayesian nonparametric models for mediation are also subject to the same RIC phenomenon as models for observational data. They make the following recommendations to combat this:
\begin{enumerate}
\item
  Add an estimate of the propensity score $\widehat \pi_i = \widehat \Pr(A_i = 1 \mid X_i = x)$ to both the outcome model and mediator model.
\item
  Add an estimate of the mediator regression function $\widehat m_{ai} = \widehat E(M_i \mid A_i = a, X_i = x)$ to the outcome regression for $a \in \{0,1\}$.
\end{enumerate}
See \citet{linero2022mediation} for an extensive discussion of why it is necessary to include these variables and a thorough simulation experiment. In principle it does not matter how $(\widehat \pi_i, \widehat m_{0i}, \widehat m_{1i})$ are obtained, aside from the fact that $\widehat \pi_i$ should depend only on $(A_i, X_i)$ and $\widehat m_{ai}$ should depend only on $(M_i, A_i, X_i)$; we use BART to estimate these quantities.


\subsection{Summarizing the Posterior}
\label{sec:summarizing}

In addition to the (conditional) average mediation effects, it is also of interest to produce interpretable summaries of the fit of the BCMF model to the data. These summaries can help identify subpopulations that respond differently to the treatment, help us interpret the impact of the effect moderators, and provide insight into BCMF's predictive process.

\citet{woody2021model} propose a general framework for posterior summarization based on projecting complex models onto interpretable surrogate models.
For example, we might project the samples of $\delta(x)$ onto an \emph{additive function} $\gamma(x) = \alpha + \sum_{j=1}^p \gamma_j(x_j)$, the idea being that if $\gamma(x)$ is a good approximation to $\delta(x)$ then we can use the interpretable structure of $\gamma(x)$ to understand how $\delta(x)$ makes predictions.

For simplicity we focus on the indirect effect $\delta(x)$ and consider two classes of summaries:
\begin{itemize}
\item
  An additive function, constructed as $\widehat \gamma = \arg \min_{\gamma} \sum_i \{\delta(X_i) - \gamma(X_i)\}^2 + q_\lambda(\gamma)$ where $\gamma(x) = \alpha + \sum_j \gamma_j(x_j)$ with each $\gamma_j(x_j)$ being a univariate spline. Here, $q_\lambda(\gamma) = \sum_j q_\lambda(\gamma_j)$ is a roughness penalty for the individual additive components. This summary can be computed by fitting a \emph{generalized additive model} \citep[GAM, see][for a review]{wood2006generalized} with $\{\delta(X_i) : i = 1,\ldots, n\}$ as the outcome and $\{X_i : i = 1,\ldots,n\}$ as the predictors.
\item
  A decision tree summary, where $\widehat \gamma$ is constructed by running the CART algorithm \citep{breiman1984classification}, treating $\{\delta(X_i) : i = 1,\ldots,n\}$ as the outcome and $\{X_i : i = 1,\ldots, n\}$ as the predictors.
\end{itemize}
CART summaries are useful for identifying subpopulations with substantially different treatment effects, while GAM summaries are useful for understanding the impact of the different predictors on the estimated effects in isolation. See Section~\ref{sec:medical} for an illustation of this approach on the MEPS dataset.

As an overall measure of the quality of the summaries we use the squared correlation between $\delta(x)$ and $\gamma(x)$ given by
\begin{align*}
  R^2 = 1 - \frac{\sum_i \{\delta(X_i) - \widehat \gamma(X_i)\}^2}{\sum_i \{\delta(X_i) - \widehat \delta\}^2}
\end{align*}
where $\widehat \delta = n^{-1} \sum_i \delta(X_i)$; \citet{woody2021model} refer to  $R^2$ as the ``summary $R^2$.''

\section{Medical Expenditure Panel Survey Data}
\label{sec:medical}

We now apply our model to a subset of the the Medical Expenditure Panel Survey (MEPS). We focus on the questions of whether (i) there is a causal effect of smoking on an individual's expected annual medical expenditures, (ii) there is evidence that the effect is entirely mediated by the effect of smoking on an health, and (iii) whether any of the proposed confounders also act as modifiers of the indirect effect. We take the outcome $Y_i$ to be the logarithm of an individual's annual net medical expenditure reported in the 2012 survey, the treatment $A_i$ to be whether an individual is a smoker (yes or no), and the mediator $M_i$ to be an ordinal measure of overall self-perceived health (1: excellent, 2: very good, 3: good, 4: fair, 5: poor).

At the outset, we note that a naive two-sample $t$-test for a difference in medical expendtures between smokers and non-smokers shows that there is strong evidence ($P$-value $< 0.0005$) that non-smokers pay \emph{more} in medical expenditures than smokers. Accordingly, it is important to control for confounders in assessing any causal relationships. Our model includes the following patient attributes as possible confounders:
\begin{itemize}
    \item \texttt{age:} Age in years.
    \item \texttt{bmi:} Body mass index, which may act as a post-treatment confounder of health and medical expenditures.
    \item \texttt{education\_level:} Education in years.
    \item \texttt{income:} Total family income per year.
    \item \texttt{poverty\_level:} Family income as percentage of the poverty line.
    \item \texttt{region:} Northeast, West, South, or Midwest.
    \item \texttt{sex:} Male or female.
    \item \texttt{marital\_status:} Married, divorced, separated, or widowed.
    \item \texttt{race:} White, Pacific Islander, Indigenous, Black, Asian, or multiple races.
    \item \texttt{seatbelt:} whether an individual wears a seatbelt in a car (always, almost always, sometimes, never, seldom, never drives/rides in a car).
\end{itemize}

\begin{figure}
  \centering
  \includegraphics[width=.9\textwidth]{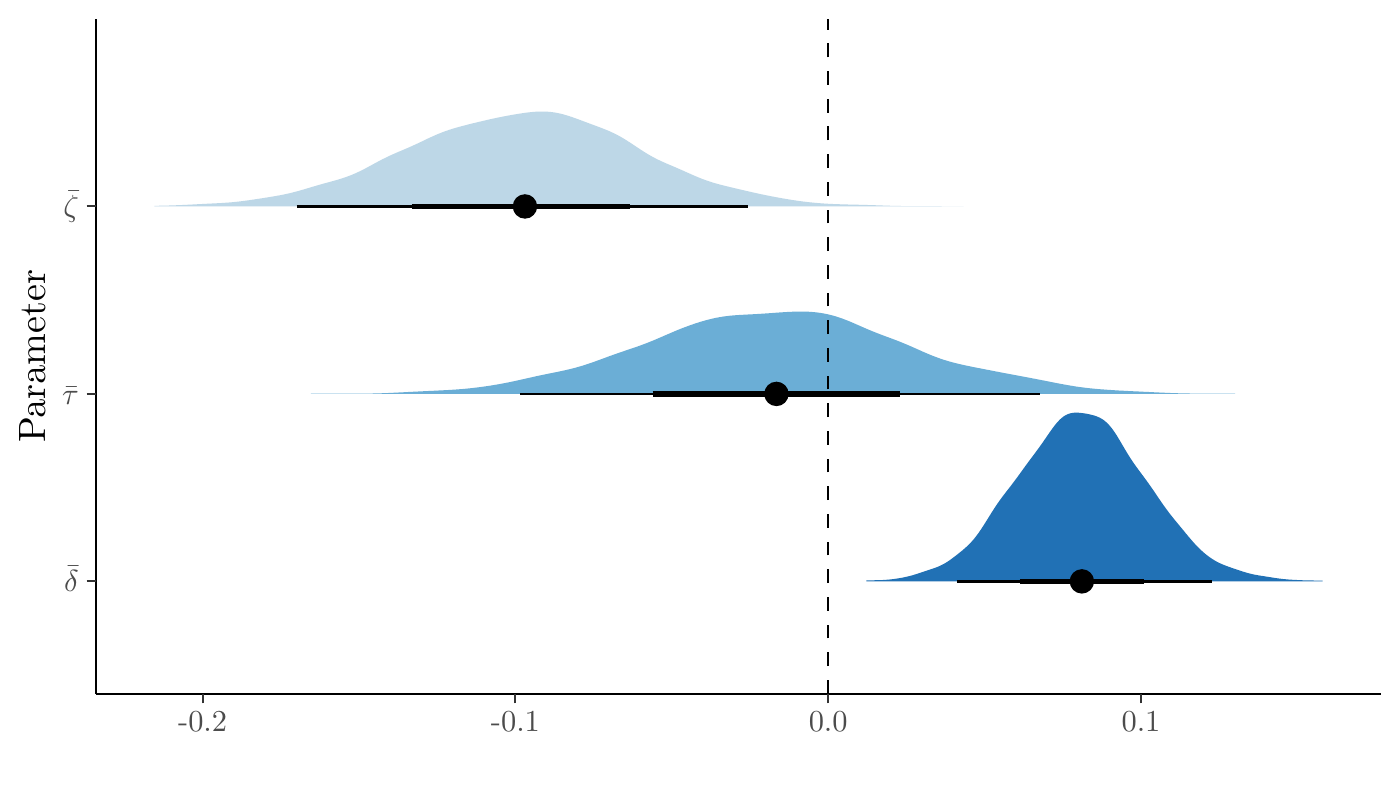}
  \caption{Posterior distribution for the average direct effect $\bar \zeta$ (top), average indirect effect $\bar \delta$ (bottom), and average total effect $\bar \tau = \bar \zeta + \bar \delta$ (middle).}
  \label{fig:avg_effects}
\end{figure}

The posterior distribution of the average direct and indirect effect is shown in Figure \ref{fig:avg_effects}. We see that, under the sequential ignorability assumption, there is evidence of both a direct and indirect effect of smoking on expenditures. Interestingly, these effects are in \emph{opposite directions} and cancel each other out to a large extent. As a result, the sign of the total effect is uncertain. This illustrates an important potential benefit of a mediation analysis: we can establish a causal relationship between smoking and medical expenditures that we could not if we restricted attention strictly to the total effect.

\begin{figure}
  \centering
  \includegraphics[width = .9\textwidth]{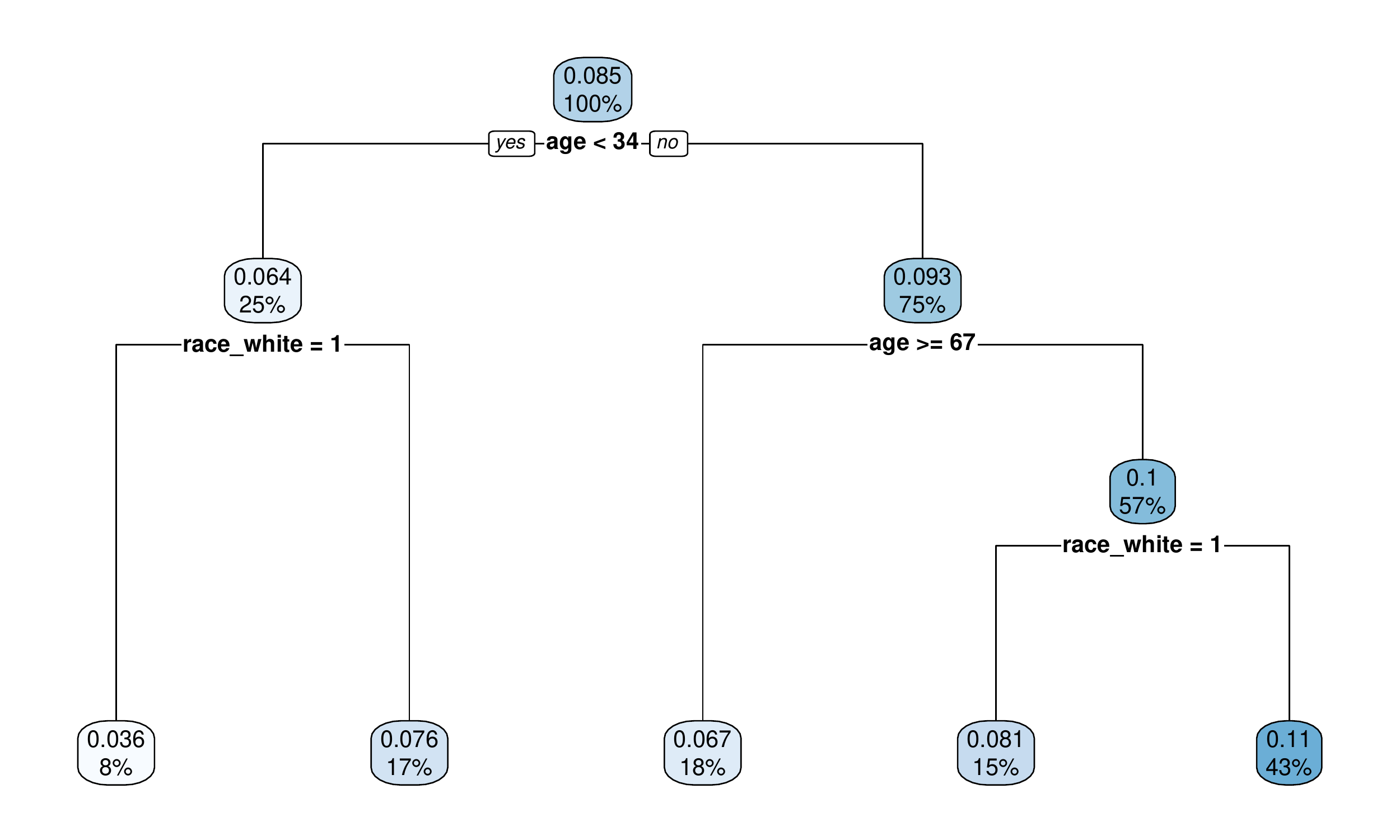}
  \caption{Posterior summarization of the indirect effect using a single regression tree.}
  \label{fig:post_tree}
\end{figure}

\begin{figure}
  \centering
  \includegraphics[width=.61\textheight]{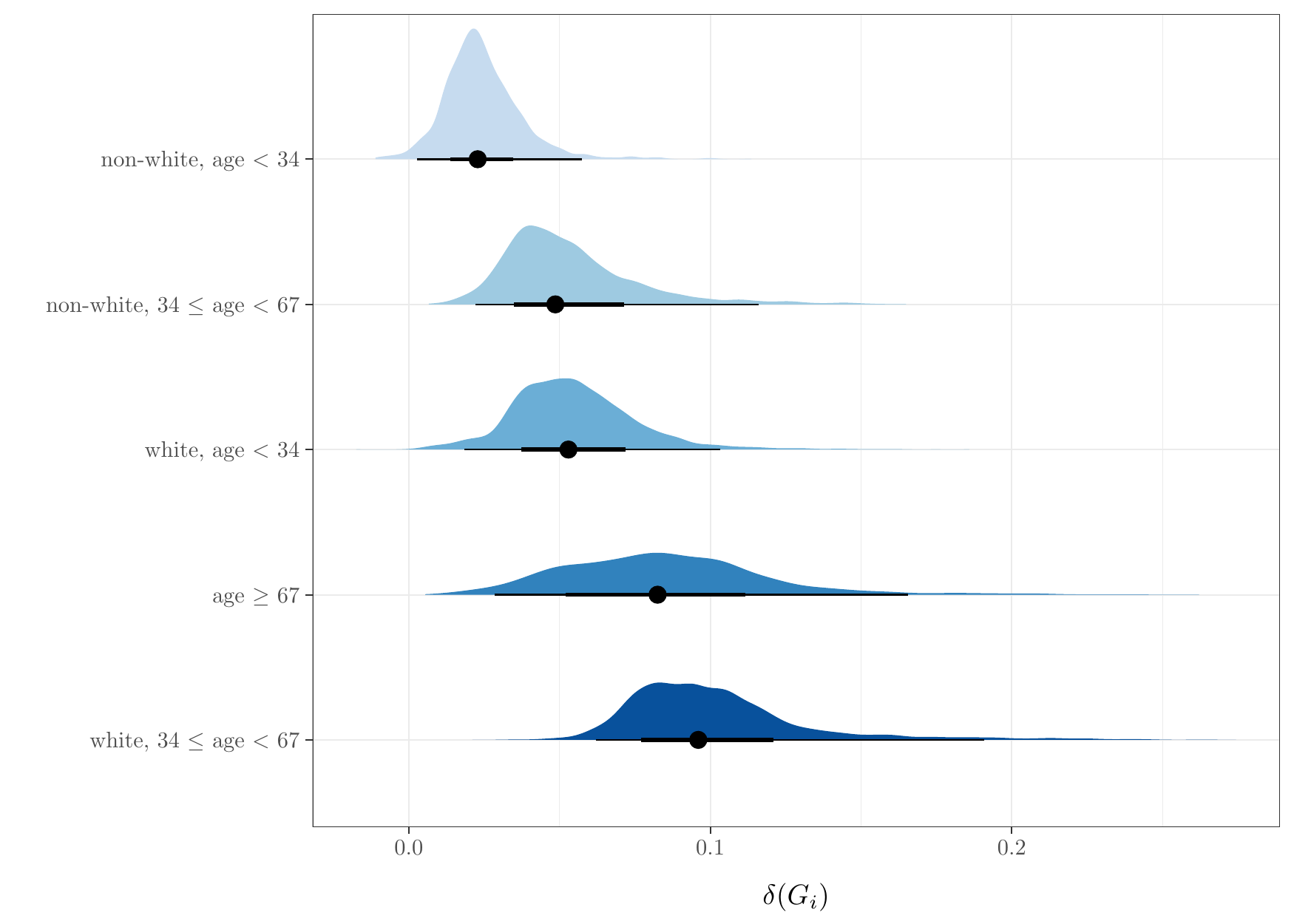}
  \caption{Posterior density for the average indirect effect within subgroups $G_1, G_2,..., G_5$ from the terminal nodes in Figure \ref{fig:post_tree}.}
  \label{fig:subgroup_indirect}
\end{figure}

\subsection{Posterior Summarization}

We use the summarization strategies outlined in Section~\ref{sec:summarizing} to interpret the model fit and better understand the covariates and interactions contributing to the heterogeneity in the indirect effect; specifically, we project the indirect effect function $\delta(x)$ onto a single regression tree and an additive function. 

We first consider a CART summary of the posterior mean of of $\delta(x)$, which was obtained on a preliminary model fit to the MEPS dataset. According to the regression tree summary in Figure~\ref{fig:post_tree}, race, age, and sex are the most significant effect modifiers for the indirect effect. Motivated by the subgroups found in Figure~\ref{fig:post_tree}, in Figure~\ref{fig:subgroup_indirect} we display the average indirect effects within various subgroups formed by age and race. We find that the largest indirect effects occur for white middle-aged individuals, while the smallest effects are for non-white young adults.

\begin{figure}
  \centering
  \includegraphics[width=.8\textwidth]{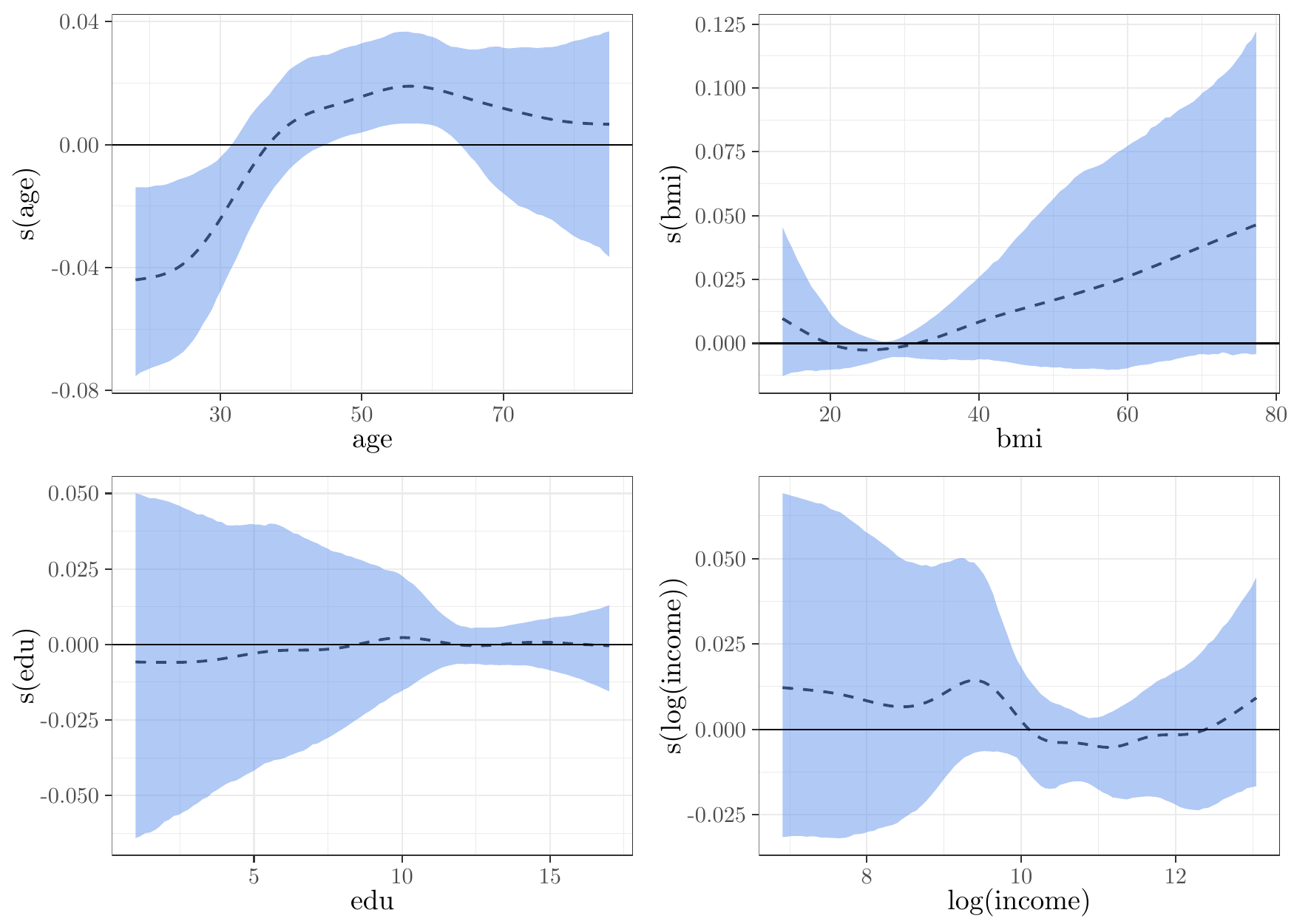}
  \caption{Posterior summarization of the indirect effect using a GAM for the continuous variables. The projection of the posterior mean is given by the dashed line while the shaded area gives a posterior 95\% credible band of the projection.}
  \label{fig:post_gam_cont}
\end{figure}
\begin{figure}
  \centering
  \includegraphics[width=1\textwidth]{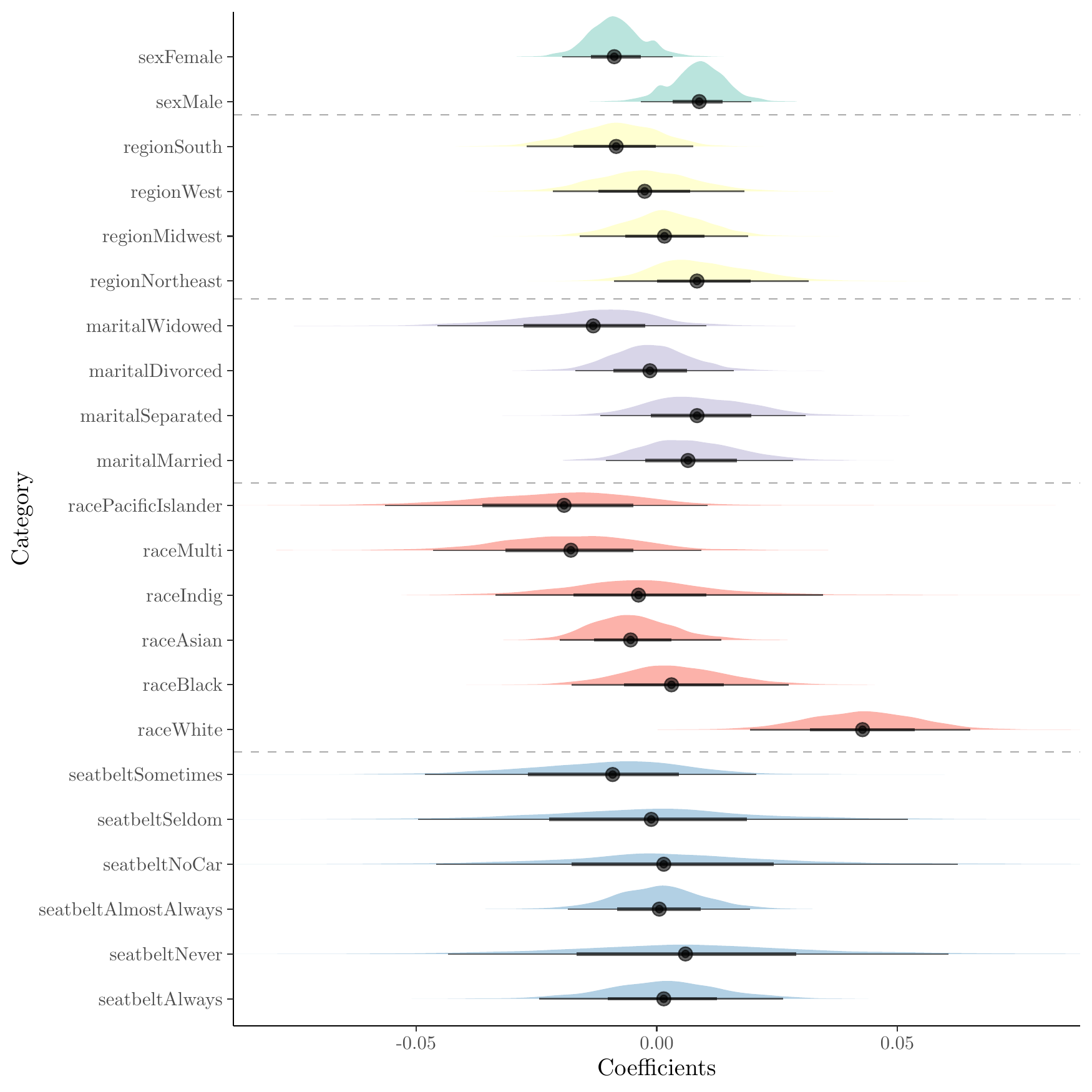}
  \caption{Posterior distributions of the impact of categorical variables on the indirect effect using the GAM projection. Solid lines give the posterior mean of the associated coefficient in the GAM model, thick bars are 66\% credible bands, thin bars are 95\% credible bands.}
  \label{fig:post_gam_cat}
\end{figure}

Figure~\ref{fig:post_gam_cont} and Figure~\ref{fig:post_gam_cat} display the results obtained from the GAM summary for continuous and discrete variables, respectively. These figures again highlight the importance of age and race as effect modifiers, indicating that older and white individuals have higher indirect effects on medical expenditures mediated by perceived health status.

\begin{figure}
  \centering
  \includegraphics[width=1\textwidth]{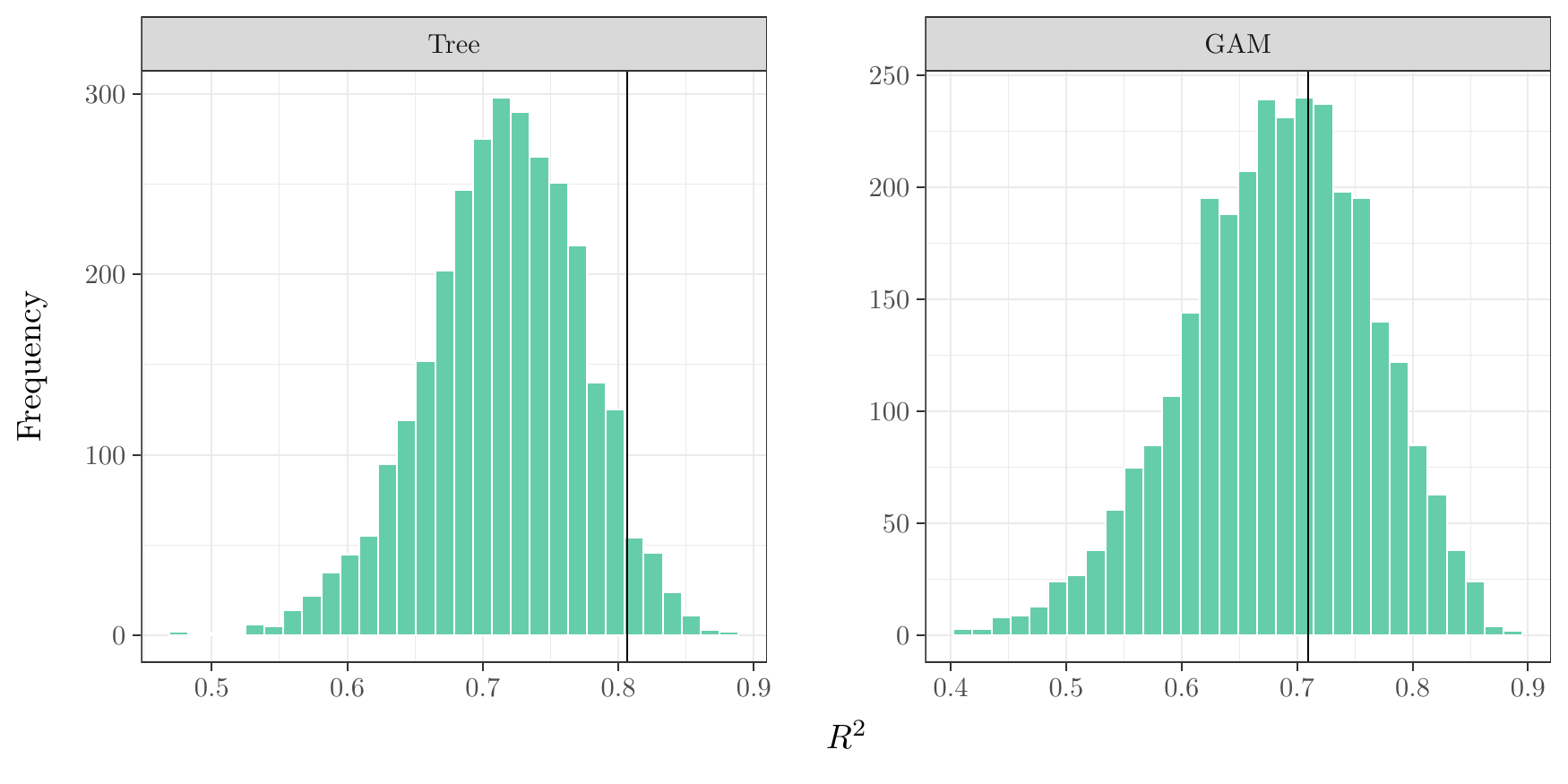}
  \caption{Posterior distribution of the summary $R^2$ for the regression tree and GAM summaries of $\delta(x)$. The black line indicates the summary $R^2$ for the posterior mean of $\delta(x)$.}
  \label{fig:R2}
\end{figure}

In summary, both the CART and GAM summaries reveal that age and race are important effect modifiers. To measure the adequacy of the summary function approximations, Figure~\ref{fig:R2} presents both the posterior distribution of $R^2$ obtained from fitting the summaries to each posterior sample of $\delta(\cdot)$ and a single $R^2$ obtained from fitting the summaries to the posterior mean of $\delta(\cdot)$. Our analysis shows that regression tree is slightly better as a summary than a GAM, suggesting that the interactions detected in Figure~\ref{fig:post_tree} provide important insight into the model's predictive process for $\delta(\cdot)$.

\subsection{Comparison of BCMF and a LSEM}
\label{sec:predictive}

To evaluate the practical usefulness of the BCMF model \eqref{eq:outcome-bart}--\eqref{eq:mediator-bart}, we compare its predictive performance to that of an LSEM with interactions between the treatment, mediator, and covariates,
\begin{align}
  \label{eq:lsem-sim}
  \begin{split}
    Y_i(a,m) &= \beta_{0Y} + X_i^T\beta_Y
               + a(\gamma_{0Y} + X_i^T\gamma_Y)
               + m(\xi_0 + X_i^T\xi) + \epsilon_i \\
    M_i(a) &= \beta_{0M} + X_i^T\beta_M +
             a(\gamma_{0M} + X_i^T\gamma_M) + \nu_i.
  \end{split}
\end{align}
This model allows for heterogeneous mediation effects, but restricts them to linear functions of the confounders. To quantify the uncertainty of the LSEM estimates, we use the residual bootstrap. By comparing the predictive performance of these two models, we can assess whether the added complexity of the BCMF model is warranted.


\begin{figure}
  \centering
  \includegraphics[width=1\textwidth]{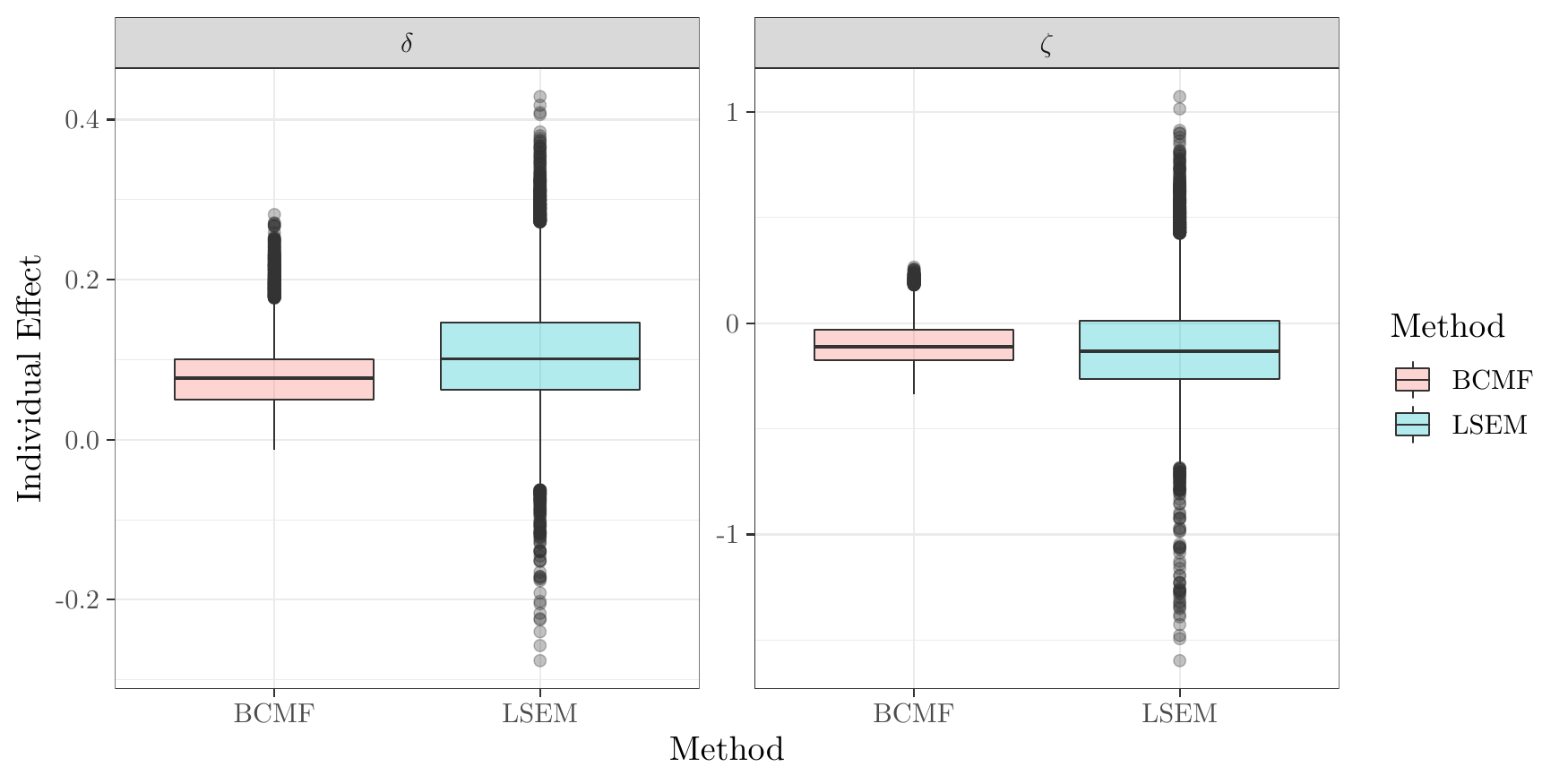}
  \includegraphics[width=1\textwidth]{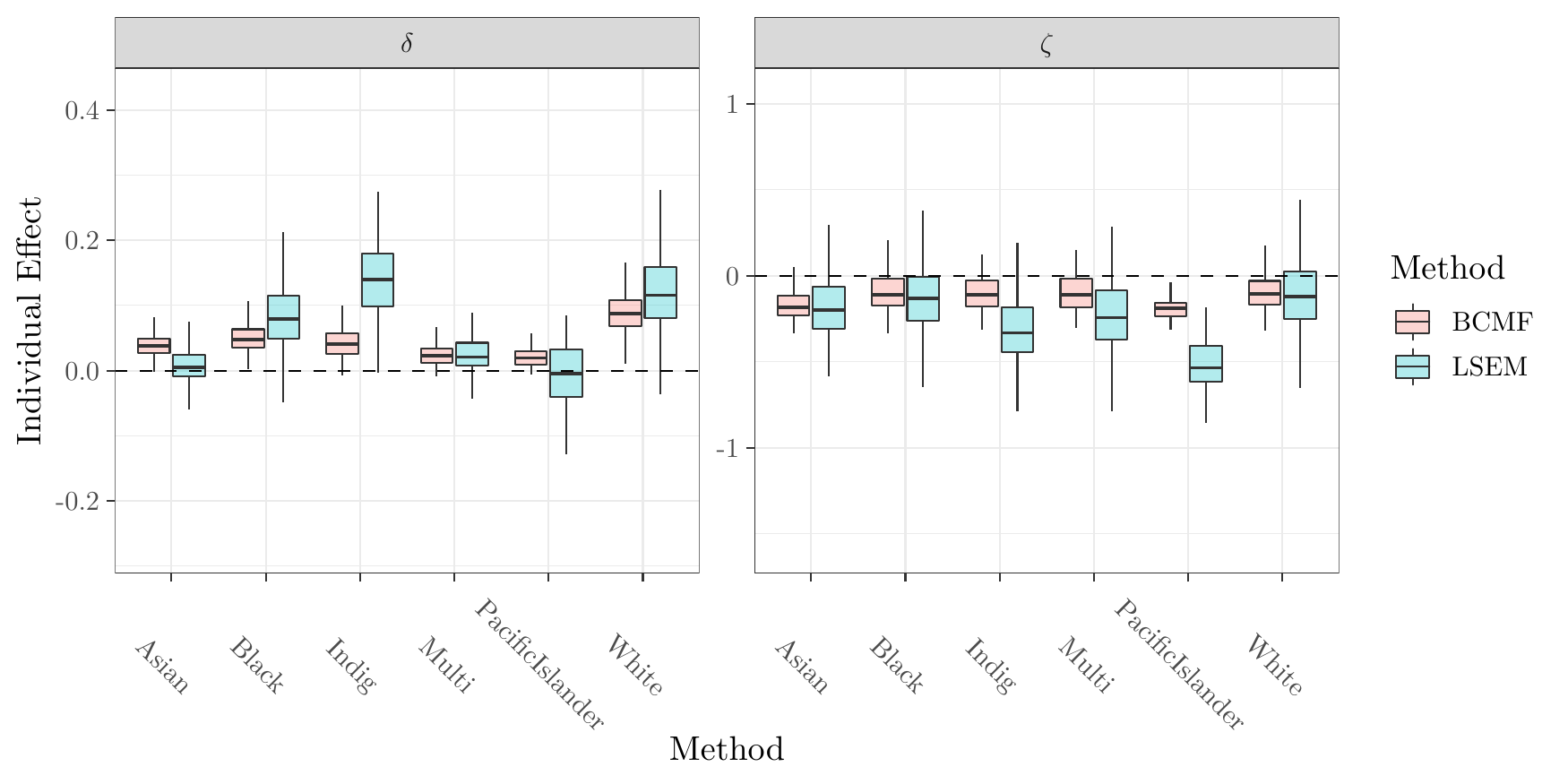}
  \caption{Top: Boxplot displaying the distribution of the estimated individual direct and indirect effects for the BCMF and LSEM models fit to the original MEPS dataset. Bottom: Boxplot displaying the distribution of the estimated individual effects for the BCMF and LSEM, stratified by race.}
  \label{fig:comparison}
\end{figure}

To understand the salient differences between the predictions made from the BCMF and LSEM, we compare the estimates of $\delta(X_i)$ and $\zeta(X_i)$ for each individual, both as a whole and stratified by race, in Figure~\ref{fig:comparison}. Figure~\ref{fig:comparison} presents the estimates of $\delta(X_i)$ and $\zeta(X_i)$ for each individual, both in aggregate and stratified by race. The effect estimates of the BCMF are substantially less heterogeneous than the LSEM, with the LSEM estimating a substantial number of both positive and negative effects for both $\delta(X_i)$ and $\zeta(X_i)$. Additionally, we see substantially less heterogeneity across race; for example, the LSEM makes counterintuitive predictions about both the direct and indirect effect of smoking within the group of Pacific Islanders. While the MEPS dataset is large, there are relatively few Pacific Islanders in the data, and in the subset we analyzed only 13 of them smoke. By applying regularization, the BCMF shrinks the direct and indirect effects within this subpopulation closer to those of the other races.

\begin{table}
    \centering
    \begin{tabular}{lrrr}
        \toprule
        Variable         & $R_{\text{test}}$ (BART) & $R_{\text{test}}$ (LSEM) & $P$-value \\
        \midrule
        \texttt{phealth} & 0.445                      & 0.419                      & $0.0001$ \\
        \texttt{log(Y)}  & 0.454                      & 0.431                      & $0.0005$ \\
        \bottomrule
    \end{tabular}
    \caption{Held-out correlation for the mediator (\texttt{phealth}) and outcome (\texttt{logY}) across all individuals for the BART and LSEM fits, and the $p$-value for a paired Wilcoxon matched pairs signed-rank test comparing the predictive performance on held-out data for the two models.}
    \label{table:heldout_rmse}
\end{table}

Next, we fit the BCMF and LSEM models to the same training set of $n = 8056$ individuals and compute predictions $(\widehat Y_{i,\lsem}, \widehat M_{i,\lsem}, \widehat M_{i,\bcmf}, \widehat Y_{i,\bcmf})$ on the test set of $n = 8057$ individuals using the fitted models. We use these predictions on the test set to evaluate the performance of the model in three ways. First, we consider the correlation between $(M_i, Y_i)$ and their predictions on the test set. Second, we perform a paired Wilcoxon signed-rank test comparing the squared difference $(Y_i - \widehat Y_{i,\lsem})^2$ to $(Y_i - \widehat Y_{i,\bcmf})^2$ (and similarly for $M_i$). Results are given in Table~\ref{table:heldout_rmse}, and we see both that the correlation is somewhat higher for the BCMF model than the LSEM model, and that the difference in performance was highly statistically significant according to the signed-rank test.

\begin{table}
  \centering
  \begin{tabular}{lrrrr}
    \toprule 
    Term                        & Estimate & Standard Error & Statistic & $P$-value \\
    \midrule
    $\widehat{M}_{\text{lsem}}$ & 0.1561   & 0.0803         & 1.9447    & 0.0518    \\
    $\widehat{M}_{\text{bcmf}}$ & 0.8477   & 0.0809         & 10.4835   & $<0.0001$ \\
    $\widehat{Y}_{\text{lsem}}$ & 0.1509   & 0.0725         & 2.0824    & 0.0373    \\
    $\widehat{Y}_{\text{bcmf}}$ & 0.9030   & 0.0747         & 12.0840   & $<0.0001$ \\
    \bottomrule
  \end{tabular}
  \caption{Coefficient estimates for the linear model that aggregates the linear and BART fits on the MEPS test data.}
  \label{tab:stacking}
\end{table}

Our third comparison considers \emph{stacking} \citep{wolpert1992stacked} the predictions of the BCMF and LSEM by fitting the linear models $Y_i = \beta_0 + \beta_1 \, \widehat Y_{i,\lsem} + \beta_2 \, \widehat Y_{i,\bcmf} + \epsilon_i$ (and similarly for $M_i$). Results of the stacking procedure are given in Table~\ref{tab:stacking}. From this fit, we see that the linear model relies much more heavily on the predictions from the BCMF than the linear model, and that the BCMF predictions are much more statistically significant than the predictions from the LSEM (in the sense that there is strong evidence that the BCMF predictor improves upon the LSEM predictor, while there is only weak evidence of the converse). Interestingly, the LSEM predictions \emph{are} found to be statistically significant, suggesting that a modification of the BCMF that also includes \emph{linear} adjustments for the confounders (i.e., includes linear terms $x^\top b$ in the functions $(\mu(x), \zeta(x), d(x), \mu_m(x), \tau_m(x))$) may improve the fit of the model.

\subsection{Simulation Study}
\label{sec:simulation}

We now conduct a simulation study to better understand the operating characteristics of the BCMF model. Our study aims to answer the following question: (i) Does the BCMF model perform better in terms of predictive accuracy in estimating the mediation effects? (ii) Do the credible intervals for $\delta(X_i)$ and $\zeta(X_i)$ attain coverage rates close to their nominal levels? (iii) Can the BCMF model estimate the effects accurately within the subgroups of the data identified by the CART summary? 

\subsubsection*{Data Generating Mechanism}

We use a data generating mechanism in which the confounders and treatment assignment are sampled direct from the MEPS dataset, while the mediator and outcome ground truths are obtained by fitting both our model and an LSEM to the data. To assess the performance of both methods, we replicated each simulation setting 200 times, with 8056 observations in the training set and 8057 in the testing set. We used the same training/testing split across all simulated datasets to evaluate the coverage probability of the confidence/credible intervals generated by each method.

A crucial difference between the LSEM and BCMF models is that the LSEM does not regularize the mediation effects. Consequently, the LSEM produces a ground truth for the mediation effects that is more heterogeneous than is expected in practice, especially for subgroups of the population with a small sample size. For example, since the MEPS dataset includes few Pacific Islanders, the LSEM's estimate of the effect of race as an effect modifier is unstable for this group. To account for this, we also consider a third ground truth that is also an LSEM but with the parameters of \eqref{eq:lsem-sim} instead estimated using the R-Learner approach of \citet{nie2021quasi}, which uses the lasso to reduce the amount of heterogeneity.

\subsubsection*{Results: Individual Effects}

We fit our model and the LSEM to each simulated dataset and measure point estimates of the effects, the limits and width of 95\% credible intervals, and whether or not the interval captures the true parameter for each replication. Using the 200 replications, we then compute the root mean square error, absolute bias, average width of the intervals, and the coverage probability.

\begin{figure}
    \centering
    \includegraphics[width=1\textwidth]{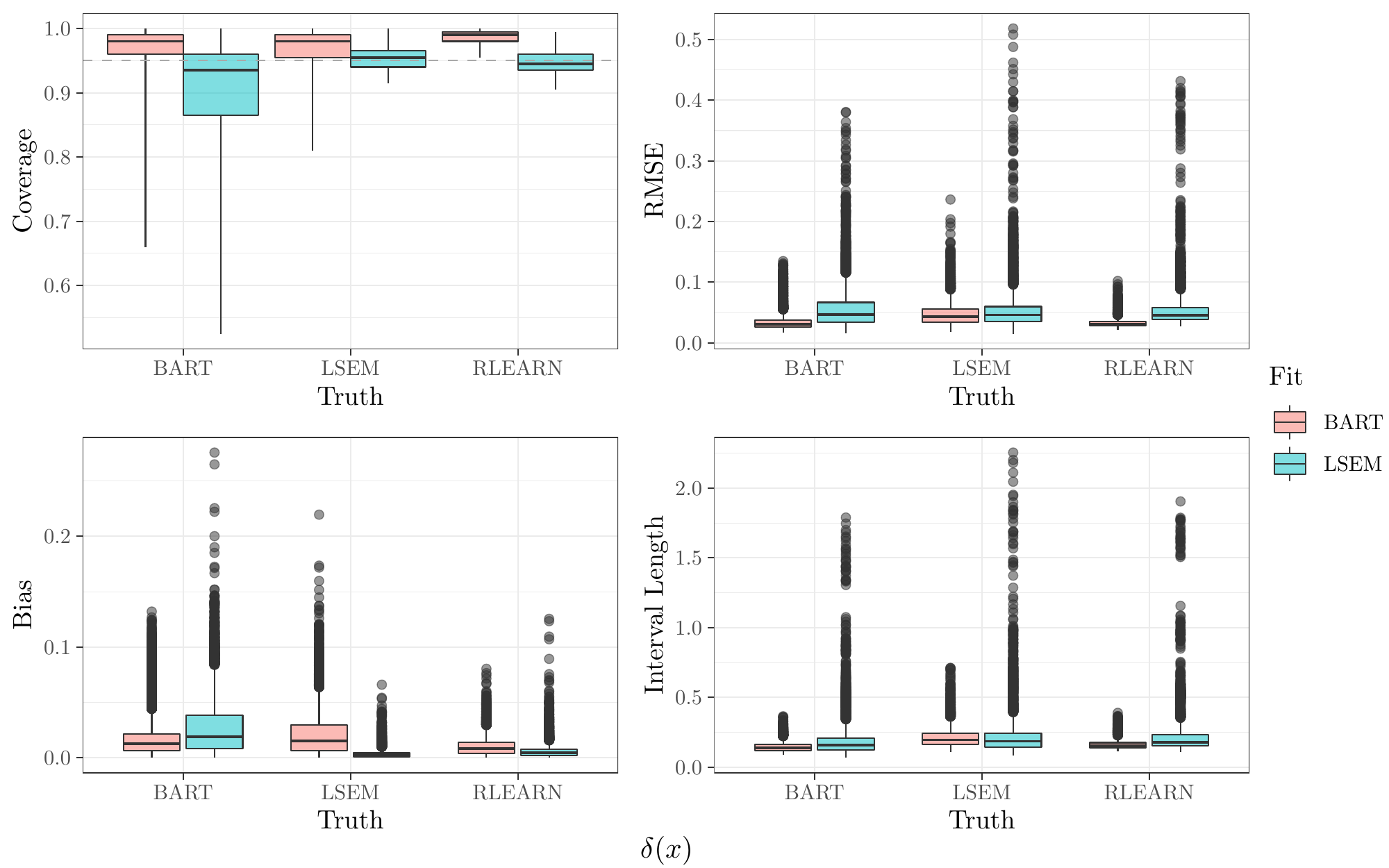}
    \caption{Individual simulation reslts for $\delta(x)$ under all combinations of fitting the BART/LSEM model under the BART/LSEM/RLEARN ground truths. Top left gives the coverage probability of nominal 95\% credible intervals among all individuals, top right gives the root mean squared error, bottom left gives the absolute bias, and bottom right gives the average interval length.}
    \label{fig:09_plot_delta}
\end{figure}

\begin{figure}
    \centering
    \includegraphics[width=1\textwidth]{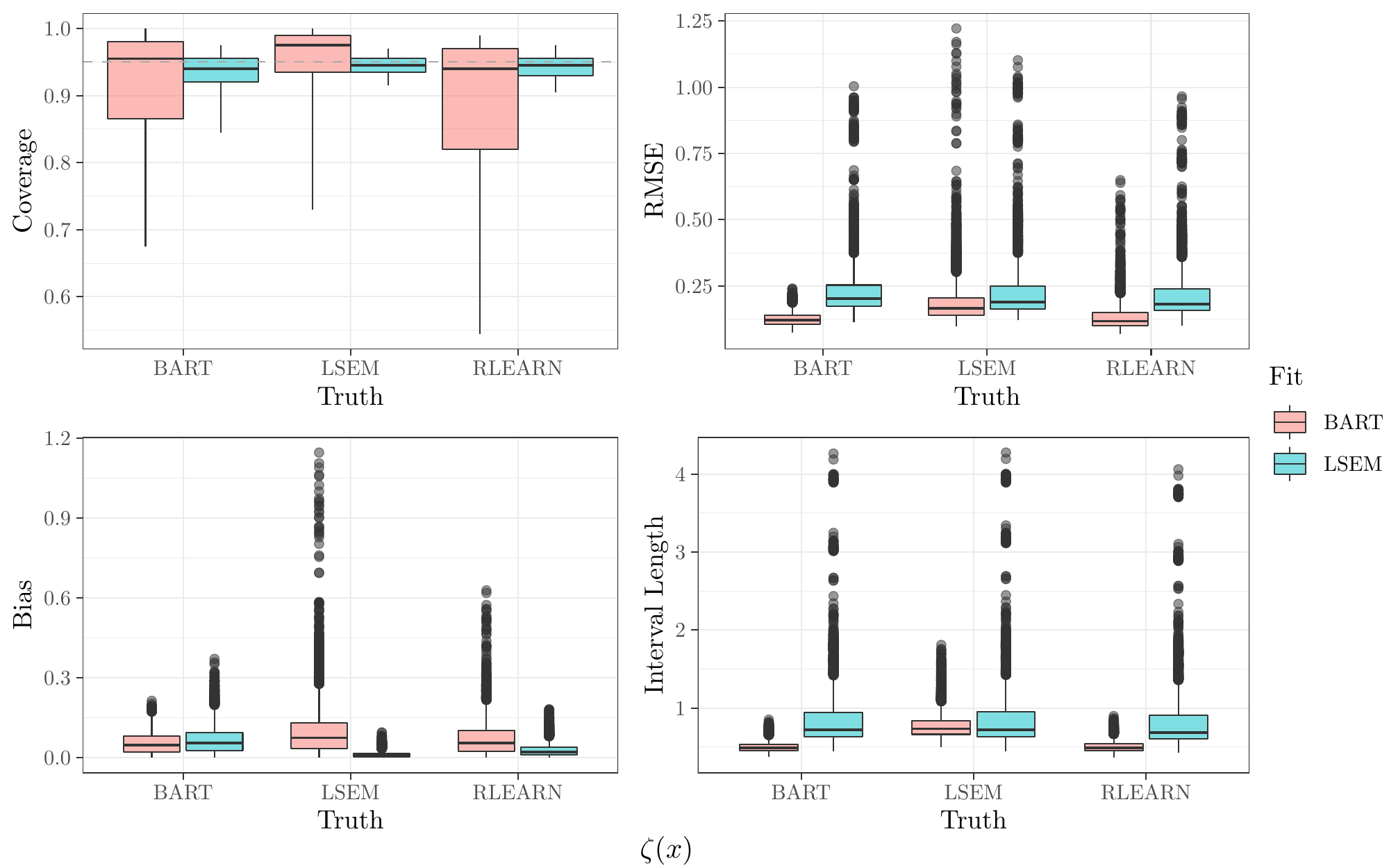}
    \caption{Individual simulation reslts for $\zeta(x)$ under all combinations of fitting the BART/LSEM model under the BART/LSEM/RLEARN ground truths. Top left gives the coverage probability of nominal 95\% credible intervals among all individuals, top right gives the root mean squared error, bottom left gives the absolute bias, and bottom right gives the average interval length.}
    \label{fig:09_plot_zeta}
\end{figure}

\begin{table}
  \begin{center}
    \begin{tabular}{llrrrr}
      \toprule
      Setting & Method & Coverage & RMSE & Bias & Length \\
      \midrule
      BART    & BART   & 0.91     & 0.12 & 0.06 & 0.50            \\
      BART    & LSEM   & 0.93     & 0.23 & 0.07 & 0.83            \\
      LSEM    & BART   & 0.94     & 0.19 & 0.10 & 0.77            \\
      LSEM    & LSEM   & 0.94     & 0.22 & 0.01 & 0.84            \\
      RLEARN  & BART   & 0.88     & 0.13 & 0.07 & 0.50            \\
      RLEARN  & LSEM   & 0.94     & 0.21 & 0.03 & 0.80            \\
      \bottomrule
    \end{tabular}
    \caption{Coverage probability, root mean square error, and absolute bias for $\zeta(X_i)$ across all individuals in the test set.} \label{table:simulation_zeta}
  \end{center}
\end{table}

\begin{table}
    \centering
    \begin{tabular}{llrrrrrr}
      \toprule
      Setting & Method & Coverage & RMSE & Bias & Interval Length \\
      \midrule
      BART    & BART   & 0.95     & 0.03 & 0.02 & 0.14            \\
      BART    & LSEM   & 0.89     & 0.06 & 0.03 & 0.19            \\
      LSEM    & BART   & 0.96     & 0.05 & 0.02 & 0.21            \\
      LSEM    & LSEM   & 0.95     & 0.05 & 0.00 & 0.22            \\
      RLEARN  & BART   & 0.99     & 0.03 & 0.01 & 0.16            \\
      RLEARN  & LSEM   & 0.95     & 0.05 & 0.01 & 0.21            \\
      \bottomrule
    \end{tabular}    
    \caption{Coverage probability, root mean square error, and absolute bias for $\delta(X_i)$ across all individuals in the test set.} \label{table:simulation_delta}
\end{table}

We present the results of our simulation study in Figure~\ref{fig:09_plot_delta} and Figure~\ref{fig:09_plot_zeta}, where we compare the BCMF and LSEM models under different combinations of ground truth and fitted models. Table~\ref{table:simulation_zeta} and Table~\ref{table:simulation_delta} summarize the results from Figure~\ref{fig:09_plot_delta} and Figure~\ref{fig:09_plot_zeta}, respectively, across all individuals in the test set. When the BCMF is fitted to the BCMF ground truth, it outperforms the LSEM in terms of achieving close to the nominal coverage on interval estimates with substantially lower interval lengths, root mean squared error, and absolute bias. Interestingly, we also observed that the BCMF model is competitive in terms of root mean squared error when the LSEM is used to generate the data. We conjecture that this is due to the fact that the data generating mechanism estimated by the LSEM fit to the original data, while still quite heterogeneous, is homogeneous enough (and the effects are small enough) that the benefits of the regularization of the BCMF outweigh the fact that the LSEM is correctly specified.
We observe similar behavior, which is even more pronounced, when the R-Learner is used to generate the ground truth.

The results for the coverage of the LSEM and BCMF models reveal some interesting results. Surprisingly, the LSEM appears to be robust in terms of coverage, although it produces much larger intervals compared to the BCMF. However, for the individual-level direct effects, the BCMF performed poorly for some individuals in achieving nominal coverage. We further investigate this behavior in the supplementary material and find that the BCMF did not attain nominal coverage for individuals with highly heterogeneous effects, i.e., those whose conditional average mediation effects $(\delta(X_i), \zeta(X_i))$ deviate greatly from the average effects $(\bar \delta, \bar \zeta)$. This behavior is expected, as the BCMF model is explicitly designed to shrink towards a structure with a small degree of heterogeneity. While this reduces the power to detect strongly heterogeneous effects, it does not inflate the Type I error in detecting heterogeneity, making the BCMF model conservative in detecting heterogeneity.


\subsubsection*{Results: Average and Subgroup Average Effects}

The BCMF also produces reliable estimates of the average mediation effects within subpopulations. We consider here both fixed and data-dependent subgroups obtained under the BCMF ground truth. The fixed subgroups are the groups identified by the terminal nodes in Figure~\ref{fig:post_tree}: age $\geq$ 67, non-white and 34 $\leq$ age $<$ 67, non-white and age $<$ 34, white and 34 $\leq$ age $<$ 67, and white and age $<$ 34. The data-determined subgroups are determined through posterior projection summarization, by fitting a tree and identifying the terminal node groups for each simulated dataset. A comparison of the inferences for the average effects under each simulation scenario is given in the supplementary material.

\begin{table}
    \centering
    \begin{tabular}{lrr}
        \toprule 
        Group                                   & Indirect Effect & Direct Effect \\
        \midrule
        $\texttt{age} \ge 67$                          & 0.99            & 0.92          \\
        $\texttt{non-white}, 34 \le \texttt{age} < 67$ & 0.96            & 0.88          \\
        $\texttt{non-white}, \texttt{age} < 34$        & 0.86            & 0.94          \\
        $\texttt{white}, 34 \le \texttt{age} < 67$     & 0.88            & 0.92          \\
        $\texttt{white}, \texttt{age} < 34$            & 0.96            & 0.96          \\
        \midrule
        Average                                        & 0.93            & 0.93          \\
        Dynamic                                        & 0.95            & 0.95          \\  
        \bottomrule
    \end{tabular}    
    \caption{Subgroup coverage probability of $\zeta_a(A_i)$ and $\delta_a(A_i)$ using the subgroups in Figure \ref{fig:post_tree}.} \label{table:fixed_sub}
\end{table}

Table~\ref{table:fixed_sub} shows the results of the simulation for both the fixed subgroups and for the data-dependent subgroups (labeled ``Dynamic''). We see that the BCMF produces intervals whose coverage is close to the nominal level, with slightly poorer results in the non-white groups. Interestingly, the coverage for data-dependent groups have \emph{higher} coverage for the credible intervals, and in fact attain exact 95\% coverage for both the direct and indirect effects. The intervals for the average effects $\bar \delta$ and $\bar \zeta$ also attain close to nominal coverage.

\section{Discussion}
\label{sec:discussion}

In this paper we introduced a Bayesian causal mediation forest (BCMF) model that can separately identify and regularize the conditional average natural direct and indirect effects using varying coefficient models. Our approach is reminiscent of LSEMs, making it easy to identify these effects as products of varying coefficients. Additionally, we demonstrate that our model produces lower prediction error than a comparable LSEM on both real and simulated MEPS data. Furthermore, we argue that our model is conservative in estimating heterogeneity since it assumes small and mostly homogeneous mediation effects. We also provide posterior summarization methods for interpreting model fit and subgroup detection.

To improve our methods and analysis, there are several directions one could take. First, we can improve the models for the outcome and mediator. For instance, log medical expenditure exhibits heteroskedasticity, with the variance of $Y_i$ and $X_i$ having a complex relationship, as demonstrated by \citet{linero2020semiparametric}. Additionally, since the mediator in this problem is ordinal, and empirically is well-approximated with a rounded normal distribution; thus, we can improve our model by using a cumulative probit model for $M_i$ rather than a normal model. The impact of using a continuous model for $M_i$ rather than an ordinal model is unclear, and warrants further investigation. 

Exclusion of individuals with no medical expenditure from the analysis (which we have done here) is problematic, as the likelihood of incurring medical expenditure is likely to be linked with smoking status. As a further improvement to our analysis, a better approach would be to use principal stratification \citep{frangakis2002principal}. This approach would estimate the causal effect of smoking on medical expenditures within the strata of individuals who incur medical expenditures, irrespective of their smoking status. In such an analysis, it is assumed that all individuals who would incur medical expenses if they did not smoke would also incur medical expenses if they did smoke. This would enable a more honest evaluation of the causal effect of smoking on medical expenditures.


Lastly, while our model performs well in terms of root mean squared error, for some individuals it does not quite reach the nominal coverage level for credible intervals. In the supplementary material, we show that our BCMF under-covers for individuals whose conditional mediation effect differs significantly from the average effects $\bar \delta$ and $\bar \zeta$. Whether this is a problem that can be fixed or simply a consequence of using a model that shrinks towards homogeneous effects warrants further investigation. Code reproducing our analysis and simulation results is available at \url{www.github.com/vcbcmf/vcbcmf}.



\bibliographystyle{apalike}
\bibliography{references.bib}

\end{document}


\date{}
\title{Supplement to Estimating Heterogeneous Causal Mediation Effects with Bayesian Decision Tree Ensembles}

\author{
  \normalsize Angela Ting and Antonio R. Linero \\
  \normalsize \emph{Department of Statistics and Data Sciences} \\
  \normalsize \emph{University of Texas at Austin}
}

\maketitle
\doublespacing

\renewcommand{\thesection}{S.\arabic{section}}
\renewcommand{\theequation}{S.\arabic{equation}}
\renewcommand{\thepage}{S.\arabic{page}}
\renewcommand{\thetheorem}{S.\arabic{theorem}}
\renewcommand{\theproposition}{S.\arabic{proposition}}
\renewcommand{\thetable}{S.\arabic{table}}
\renewcommand{\thefigure}{S.\arabic{figure}}
\renewcommand{\thelemma}{S.\arabic{lemma}}
\thispagestyle{fancy}
\pagestyle{fancy}
\rhead{\textbf{Not for Publication Supplementary Material}}

\section{Bayesian Backfitting Algorithm}

We now present a Bayesian backfitting algorithm for fitting the BCMF model of Section~\ref{sec:bart_mediation}. From Section \ref{sec:bart}, BART induces a posterior on $(T_1, M_1),..., (T_m, M_m)$. We use a Bayesian backfitting algorithm \citep{chipman2010bart} to update each $(T_j, M_j)$, which is based on treating the partial residuals excluding the $j^{\text{th}}$ tree as the response. Let $T_{(j)}$ denote the set of all trees in the sum except tree $T_j$. Similarly, let $M_{(j)}$ denote the set of all terminal nodes except $M_j$. Then, for a sum of trees model $Y_i = \sum_{j=1}^m g(X_i; T_j, M_j) + \epsilon_i$, the vector of partial residuals which excludes the $j$th tree is
\begin{align*}
  R_{ij} \equiv Y_i - \sum_{k \neq j} g(X_i; T_k, M_k).
\end{align*}
$R_{ij}$ can be used to update $(T_j, M_j)$ using the single-tree Metropolis-Hastings algorithm described by \citet{chipman1998bayesian} with $R_{ij}$ now taking the place of the data $Y_i$ since $R_{ij} = g(X_i; T_j, M_j) + \epsilon$ corresponds to a single-tree model.

In our model, $\mu, \zeta, d, \mu_m$, and $\tau_m$ are given BART priors. The mediation model parameters $\mu_m$ and $\tau_m$ can be updated using the Bayesian backfitting algorithm given in \citet{hahn2020bayesian}. To handle the outcome model parameters, let
\begin{align*}
    \mu(x) &= \sum_j g(x; T_j^{\mu}, M_j^{\mu}) \\
    \zeta(x) &= \sum_j g(x; T_j^{\zeta}, M_j^{\zeta}) \\
    d(x) &= \sum_j g(x; T_j^{d}, M_j^{d}).
\end{align*}
Then we define the partial residual for $\mu$ as
\begin{align*}
    R_{i}^{\mu} &= Y_i - A_i \, \zeta(X_i) - M_i \, d(X_i) - \sum_{k \neq j} g(X_i; T_k^{\mu}, M_k^{\mu}) \\
    &=  g(X_i; T_j^{\mu}, M_j^{\mu}) + \epsilon_i
\end{align*}
where $\epsilon_i\sim N(0, \sigma^2)$. Therefore, as in \citet{chipman1998bayesian}, $(T_j^{\mu_Y}, M_j^{\mu_Y})$ can be updated with BART using the above residual as the new response. Similarly, $(T_j^{\zeta}, M_j^{\zeta})$ can be updated with BART using the following partial residual
\begin{align*}
  R_i^\zeta
  &= Y_i - \mu(X_i) - M_i \, d(X_i)
    - A_i \, \sum_{k \ne j} g(X_i; T_k^\zeta, M_k^\zeta) \\
  &= A_i \, g(X_i ; T_j^\zeta, M_j^\zeta) + \epsilon_i.
\end{align*}
Because $A_i$ is binary, this implies that $(T_j^\zeta, M_j^\zeta)$ can be updated using the Bayesian CART algorithm proposed by \citet{chipman1998bayesian} applied to the observations $R_i^\zeta$ such that $A_i = 1$. Finally, let the partial residuals for $d$ be given by
\begin{align*}
  R_i^{d} &= \frac{Y_i - \mu(X_i) - A_i\zeta(X_i)}{M_i}
            - \sum_{k \neq j} g(X_i; T_k^{d}, M_k^{d}) \\
    &= g(X_i; T_j^d, M_j^d) + \eta_i
\end{align*}
where $\eta_i = \frac{\epsilon_i}{M_i} \sim N\left( 0, \frac{\sigma^2}{M_i^2} \right)$. Hence, $(T_j^d, M_j^d)$ can be updated using an update for the heteroskedastic variant of the Bayesian CART with known weights; an algorithm for performing an update for this model is given, for example, by \citet{pratola2020heteroscedastic}.

\section{Additional Simulation Results}

\begin{table}[p]
  \centering
  \begin{tabular}{llrrrr}
    \toprule
    Setting & Method & Coverage & RMSE & Bias & Length \\
    \midrule
    BART & BART & 0.96 & 0.01 & 0.00 & 0.05\\
    BART & LSEM & 0.96 & 0.01 & 0.01 & 0.05\\
    LSEM & BART & 0.98 & 0.01 & 0.00 & 0.06\\
    LSEM & LSEM & 0.96 & 0.01 & 0.00 & 0.06\\
    RLEARN & BART & 0.98 & 0.01 & 0.00 & 0.06\\
    RLEARN & LSEM & 0.96 & 0.01 & 0.00 & 0.05\\
    \bottomrule
  \end{tabular}
  \caption{Coverage probability, root mean square error, and absolute bias for $\bar \delta$.} \label{table:simulation_delta_avg}
\end{table}

\begin{table}[p]
  \begin{center}
    \begin{tabular}{llrrrr}
      \toprule
      Setting & Method & Coverage & RMSE & Bias & Length \\
      \midrule
      BART & BART & 0.90 & 0.06 & 0.02 & 0.21\\
      BART & LSEM & 0.96 & 0.06 & 0.01 & 0.21\\
      LSEM & BART & 0.97 & 0.06 & 0.02 & 0.24\\
      LSEM & LSEM & 0.96 & 0.05 & 0.00 & 0.22\\
      RLEARN & BART & 0.90 & 0.05 & 0.02 & 0.20\\
      RLEARN & LSEM & 0.94 & 0.05 & 0.01 & 0.20\\
      \bottomrule
    \end{tabular}
    \caption{Coverage probability, root mean square error, and bias for $\bar \zeta$.} \label{table:simulation_zeta_avg}
  \end{center}
\end{table}

In this section we provide additional results from the simulation study of Section~\ref{sec:simulation}. In Table~\ref{table:simulation_delta_avg} and Table~\ref{table:simulation_zeta_avg} we give the coverage for the parameters $\bar \delta$ and $\bar \zeta$ respectively under all combinations of ground truth (BART, LSEM, and the R-Learner) and estimation method (BART or LSEM). We see in Table~\ref{table:simulation_delta_avg} that the BCMF and LSEM models perform essentially the same, with BART having marginally higher coverage and comperable RMSE, absolute bias, and interval length. In Table~\ref{table:simulation_zeta_avg} we find that the BCMF produces slightly more biased estimates of the direct effect, leading to lower coverage overall.

\begin{figure}
  \centering
  \includegraphics[height= 1\textwidth]{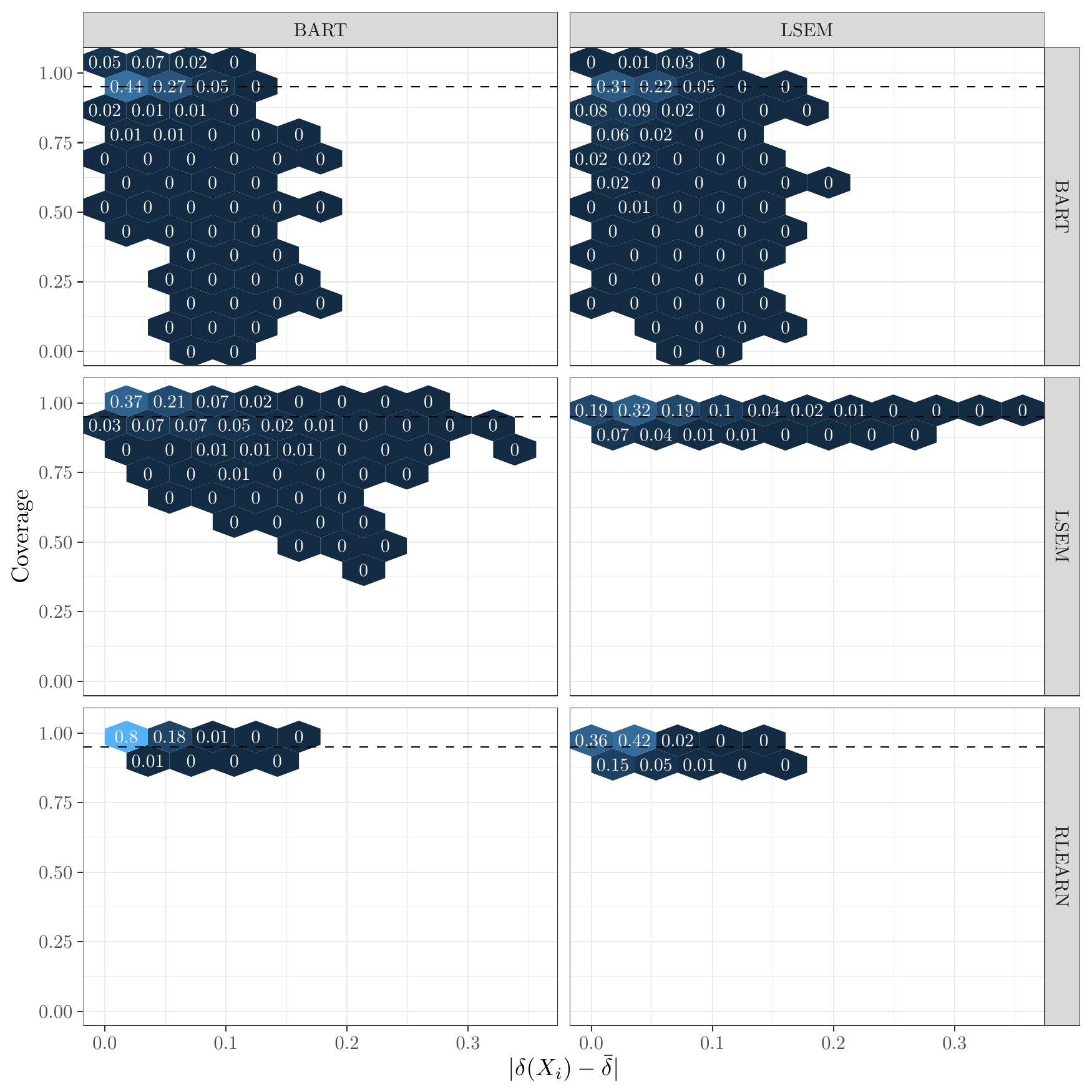}
  \caption{Coverage of the BCMF (left) and LSEM (right) models for $\delta(X_i)$ as a function of $|\delta(X_i) - \bar \delta|$, displayed as a two-dimensional hexogonal histogram. The proportion of individuals falling in each bin is given by the bin label.}
  \label{fig:coverage_by_size_delta}
\end{figure}

\begin{figure}
  \centering
  \includegraphics[height= 1\textwidth]{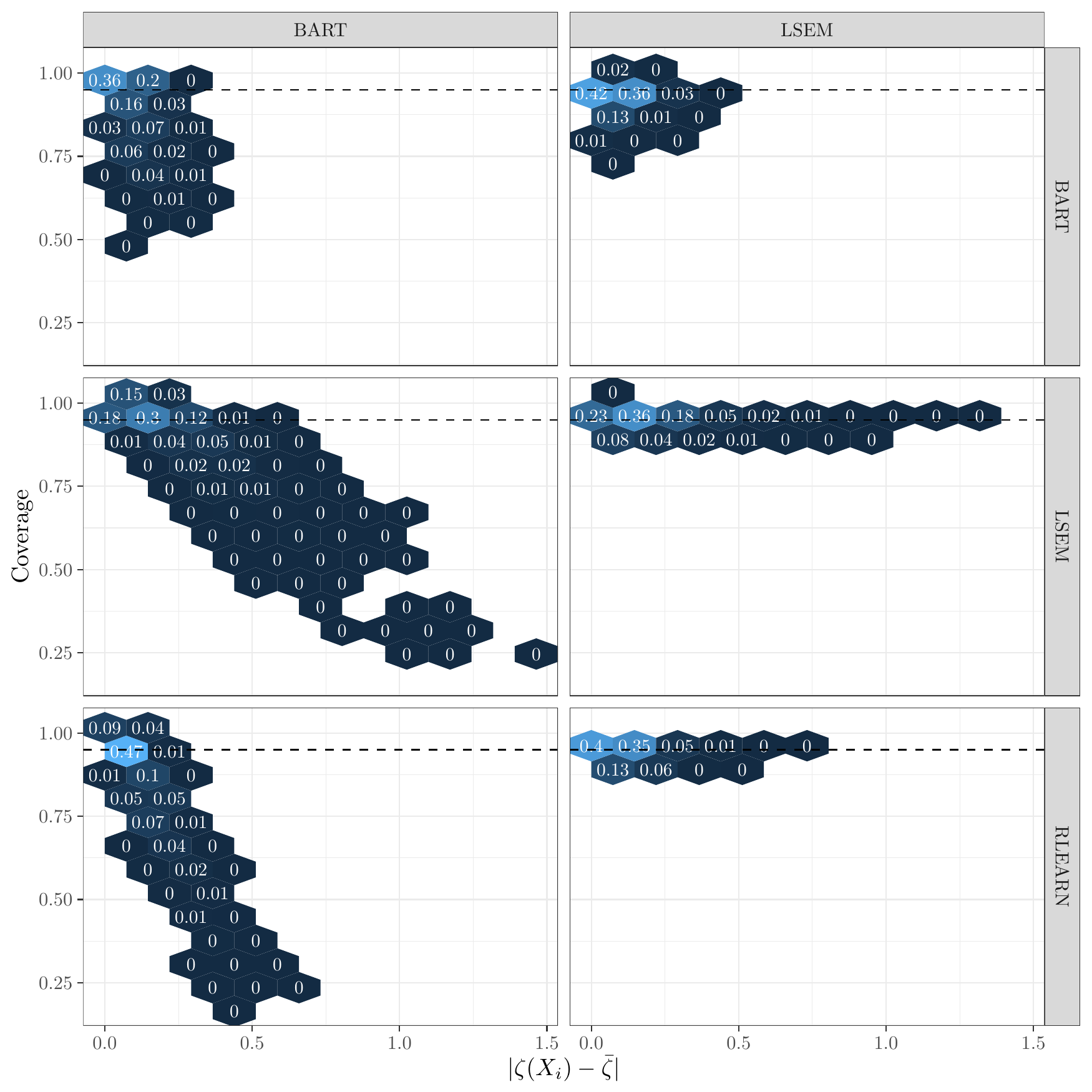}
  \caption{Coverage of the BCMF (left) and LSEM (right) models for $\zeta(X_i)$ as a function of $|\zeta(X_i) - \bar \zeta|$, displayed as a two-dimensional hexogonal histogram. The proportion of individuals falling in each bin is given by the bin label.}
  \label{fig:coverage_by_size_zeta}
\end{figure}

In Figure~\ref{fig:coverage_by_size_delta} and Figure~\ref{fig:coverage_by_size_zeta} we validate the claim in the manuscript that the coverage of the BCMF tends to be worse for individuals whose conditional average effects differ greatly from the average effect. Because the LSEM does not shrink towards homogeneity, we see that the coverage for both the direct and indirect effects is relatively constant as a function of $|\zeta(X_i) - \bar \zeta|$ and $|\delta(X_i) - \bar \delta|$, while the BCMF tends to have poorer coverage as these quantities increase.

\bibliographystyle{apalike}
\bibliography{references.bib}